\documentclass[10pt,conference]{IEEEtran}

\usepackage{float}
\usepackage{xcolor}
\usepackage{multirow}
\usepackage{paralist}
\usepackage{booktabs}
\usepackage{xspace}
\usepackage{subcaption}
\usepackage{balance} 
\usepackage{graphicx}
\usepackage{amsfonts}

\newcommand{\rev}[1]{{\textcolor{black}{{#1}}}}
\newcommand{\revicde}[1]{{\textcolor{blue}{{#1}}}}

\definecolor{black}{rgb}{0,0,0}
\definecolor{grey}{rgb}{0.8,0.8,0.8}
\definecolor{red}{rgb}{1,0,0}
\definecolor{green}{rgb}{0,1,0}
\definecolor{darkgreen}{rgb}{0,0.5,0}
\definecolor{darkpurple}{rgb}{0.5,0,0.5}
\definecolor{darkdarkpurple}{rgb}{0.3,0,0.3}
\definecolor{blue}{rgb}{0,0,1}
\definecolor{shadegreen}{rgb}{0.95,1,0.95}
\definecolor{shadeblue}{rgb}{0.95,0.95,1}
\definecolor{shadered}{rgb}{1,0.85,0.85}
\definecolor{shadegrey}{rgb}{0.85,0.85,0.85}
\definecolor{oddRowGrey}{rgb}{0.80,0.80,0.80}
\definecolor{evenRowGrey}{rgb}{0.85,0.85,0.85}

\definecolor{ForestGreen}{rgb}{0.0, 0.66, 0.47}
\definecolor{RubineRed}{rgb}{1.0, 0.0, 0.31}

\newcommand{\green}[1]{{\textcolor{ForestGreen}{{#1}}}}

\newcommand{\red}[1]{{\textcolor{RubineRed}{{#1}}}}

\newtheorem{definition}{Definition}

\newtheorem{example}{Example}

\newcommand{\system}{\ensuremath{\mathsf{Sudowoodo}}\xspace}

\newcommand{\di}{{DI\&P}\xspace}
\newcommand{\ditto}{\textsf{Ditto}\xspace}

\newcommand{\dm}{\textsf{DeepMatcher}\xspace}
\newcommand{\zeroer}{\textsf{ZeroER}\xspace}
\newcommand{\autofj}{\textsf{Auto-FuzzyJoin}\xspace}
\newcommand{\baran}{\textsf{Baran}\xspace}
\newcommand{\sato}{\textsf{Sato}\xspace}
\newcommand{\sherlock}{\textsf{Sherlock}\xspace}
\newcommand{\dbl}{\textsf{DL-Block}\xspace}
\newcommand{\rot}{\textsf{Rotom}\xspace}

\usepackage[shortlabels]{enumitem}
\setlist[itemize]{leftmargin=*}
\setlist[enumerate]{leftmargin=*}

\usepackage[linesnumbered,ruled,vlined]{algorithm2e}

\usepackage{dsfont}

\usepackage{amsmath}
\DeclareMathOperator*{\argmax}{arg\,max}

\usepackage{multirow}
\usepackage[font=small,skip=3pt]{caption}

\AtBeginDocument{%
  \providecommand\BibTeX{{%
    \normalfont B\kern-0.5em{\scshape i\kern-0.25em b}\kern-0.8em\TeX}}}

\begin{document}
\title{\system: Contrastive Self-supervised Learning for Multi-purpose Data Integration and Preparation}

\author{\IEEEauthorblockN{Runhui Wang}
\IEEEauthorblockA{\textit{Computer Science Department} \\
\textit{Rutgers University}\\
New Brunswick, United States \\
runhui.wang@rutgers.edu}
\and
\IEEEauthorblockN{Yuliang Li}
\IEEEauthorblockA{ 
\textit{Megagon Labs}\\
Mountain View, United States \\
yuliang@megagon.ai}
\and
\IEEEauthorblockN{Jin Wang}
\IEEEauthorblockA{
\textit{Megagon Labs}\\
Mountain View, United States \\
jin@megagon.ai}
}

\maketitle

\begin{abstract}

Machine learning (ML) is playing an increasingly important role in data management tasks, particularly in Data Integration and Preparation (\di).
The success of ML-based approaches, however, heavily relies on the availability of large-scale, high-quality labeled datasets
for different tasks. 
Moreover, the wide variety of \di tasks and pipelines
oftentimes requires customizing ML solutions at a significant cost for model engineering and
experimentation.
These factors inevitably hold back the adoption of ML-based approaches to new domains and tasks.

In this paper, we propose \system, a multi-purpose \di framework 
based on contrastive representation learning. 
\system features a unified, matching-based problem definition
capturing a wide range of \di tasks including 
Entity Matching (EM) in data integration, error correction in data cleaning, semantic type detection in data discovery, and more.
Contrastive learning enables \system to 
learn similarity-aware data representations from a large corpus
of data items (e.g., entity entries, table columns) without using any labels. 
The learned representations can later be either directly used
or facilitate fine-tuning with only a few labels
to support different \di tasks.
Our experiment results show that \system achieves multiple
state-of-the-art results on different levels of supervision
and outperforms previous best 
specialized blocking or matching solutions for EM.
\system also achieves promising results in data cleaning and 
column matching tasks
showing its versatility in \di applications. 
\end{abstract}


\section{Introduction}\label{sec:intro}

Machine learning, particularly deep learning (DL), is playing an increasingly important
role in almost all fields in computer science including data management~\cite{DBLP:conf/naacl/SocherM13,DBLP:journals/sigmod/0059Z0JOT16,DBLP:journals/ar/PiersonG17,DBLP:journals/bib/BerrarD21}.
More recently, this trend has extended to Data Integration and Preparation (\di)~\cite{DBLP:books/daglib/0029346,DBLP:journals/debu/ChenGHTD18,DBLP:journals/sigmod/HameedN20,DBLP:journals/pvldb/TangFLTDLMO21}
with learning-based solutions making promising progress and advancing the state-of-the-art
in multiple tasks, including entity matching~\cite{DBLP:conf/sigmod/MudgalLRDPKDAR18,ditto2021}, 
data cleaning~\cite{DBLP:conf/sigmod/MahdaviAFMOS019,DBLP:conf/cidr/MahdaviA21,DBLP:journals/pvldb/MahdaviA20,DBLP:conf/sigmod/HeidariMIR19},
and table annotation~\cite{DBLP:journals/pvldb/ZhangSLHDT20,DBLP:journals/pvldb/DengSL0020,DBLP:conf/www/WangSLHDJ21}. 
For example, Entity Matching (EM) solutions based on pre-trained language models such as BERT~\cite{DBLP:conf/naacl/DevlinCLT19}
achieve new state-of-the-art results across multiple matching tasks~\cite{DBLP:conf/edbt/BrunnerS20,ditto2021}.
While shown to be effective, as we describe next, 
existing learning-based solutions for \di
suffer from two major challenges making them less attractive in practice.

\smallskip
\noindent
\textbf{Challenge 1: Label Requirement. }
The success of ML-based approaches comes at the cost of
creating large-scale, high-quality annotated datasets. 
For example, in Entity Matching, learning-based methods typically require
around 10k or more labeled entity pairs to achieve the best pairwise matching performance~\cite{DBLP:conf/sigmod/MudgalLRDPKDAR18,ditto2021}.
Creating high-quality datasets of such cardinality can be quite expensive considering the required human efforts, 
which make ML-based solutions less accessible to new \di tasks and domains.

\smallskip
\noindent
\textbf{Challenge 2: Task Variety. } 
There is a wide scope of \di tasks requiring ML models
for different problem formulations. This means that oftentimes,
practitioners have to build a specialized 
ML solution for each task,
resulting in extra costs of model engineering. 
For example, Entity Matching typically consists of two tasks,
\emph{blocking} and \emph{matching}~\cite{DBLP:journals/csur/PapadakisSTP20},
with the goal of (1) filtering out candidate matching pairs and (2) performing the pairwise comparison, respectively.
The two tasks are closely related, but because of their 
different problem formulations, 
existing studies typically propose specialized solutions 
~\cite{DBLP:journals/pvldb/EbraheemTJOT18,DBLP:conf/sigmod/MudgalLRDPKDAR18,DBLP:conf/edbt/BrunnerS20,DBLP:journals/pvldb/ThirumuruganathanLTOGPFD21,ditto2021} that require separate
modeling, data annotation, and experimentation efforts
for each task.
This has also been the case for other \di tasks
in data cleaning
and data discovery~\cite{DBLP:conf/sigmod/MahdaviAFMOS019,DBLP:conf/cidr/MahdaviA21,DBLP:journals/pvldb/MahdaviA20,DBLP:conf/sigmod/HeidariMIR19,DBLP:journals/pvldb/ZhangSLHDT20,DBLP:journals/pvldb/DengSL0020,DBLP:conf/www/WangSLHDJ21}.

\begin{figure}[ht]
    \centering
    \includegraphics[width=0.48\textwidth]{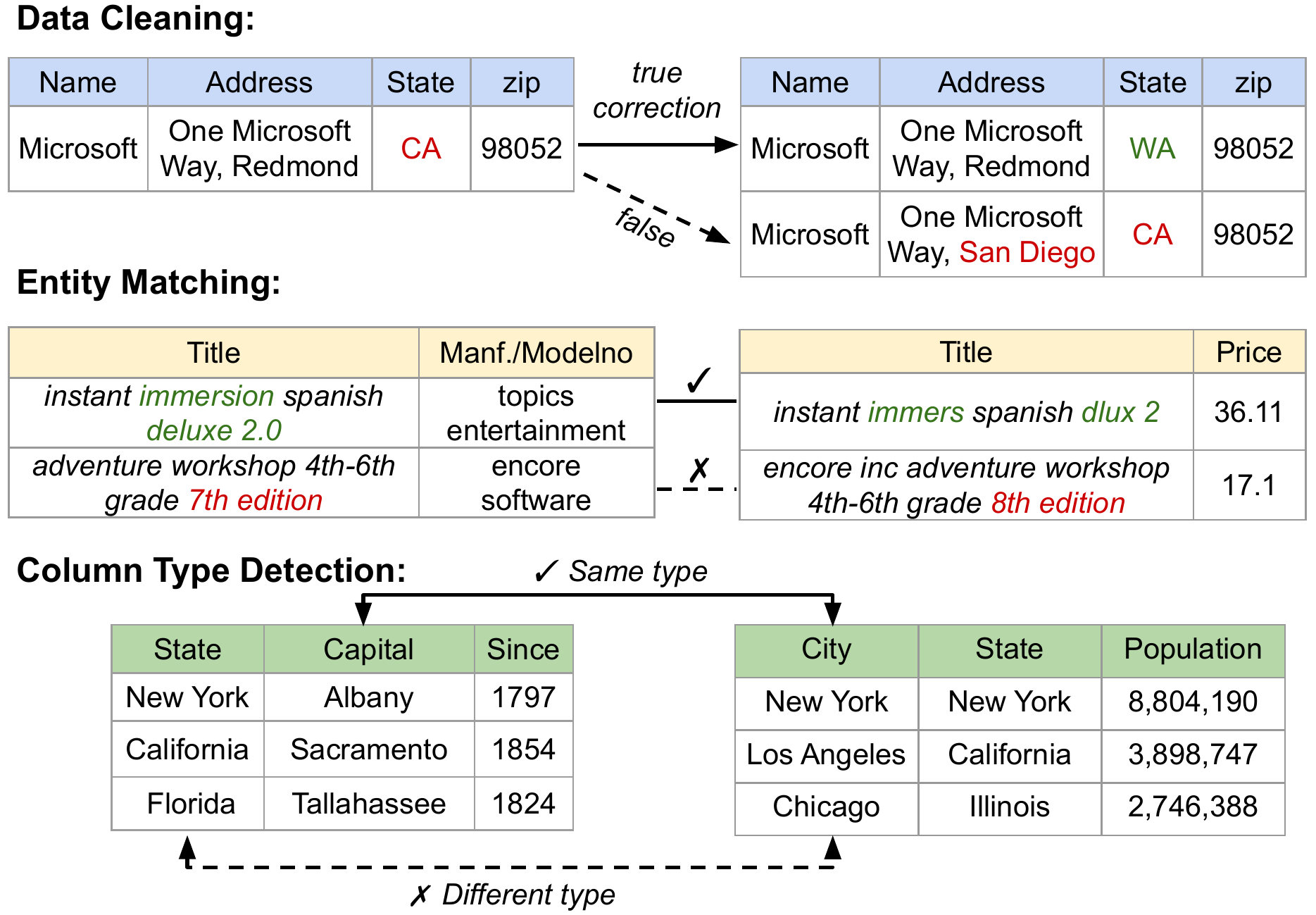}
    \caption{ \system supports any \di tasks that can be formulated as a general form of matching task. }
    \label{fig:definition}
    \vspace{-5mm}
\end{figure}

To address these two challenges, we proposed \system, 
a multi-purpose \di framework based on
\emph{contrastive representation learning}~\cite{DBLP:conf/icml/ChenK0H20}.
\system features a generic \di problem definition of data matching:
given a collection of data items (e.g., entity entries, 
table columns, cell values), identify all pairs of 
compatible items. 
This generic definition allows
\system to support a wide range of tasks
in data integration, cleaning, and discovery.
As Figure \ref{fig:definition} illustrates,
EM can be formulated as a task of matching entity entries;
error correction in data cleaning 
\cite{DBLP:conf/cidr/MahdaviA21,DBLP:journals/pvldb/MahdaviA20,DBLP:conf/sigmod/HeidariMIR19} can be formulated as matching erroneous cell values with candidate corrections;
and column matching, which is widely used in data discovery
\cite{DBLP:conf/kdd/HulsebosHBZSKDH19,DBLP:journals/pvldb/ZhangSLHDT20},
can be formulated
as matching table columns that have the same semantic type. 
The matching criteria are also customizable w.r.t the specific application.
In this work, we consider any criteria that can be cast into a general binary relation,
such as whether two entity entries refer to the same real-world entity for EM.

\system addresses the label requirement challenge
by leveraging contrastive learning to 
learn a data representation model
from a large collection of unlabeled data items,
such as corpora of entity entries or table columns.
Unlike that of language models such as BERT~\cite{DBLP:conf/naacl/DevlinCLT19}, 
the effectiveness of \system's pre-training process relies on
its contrastive learning objective instead of learning linguistic characteristics. 
The contrastive objective allows \system to learn 
how to distinguish pairs of similar data items from dissimilar ones that are likely to be distinct.
In this way, without using any labels,
\system learns meaningful representations where similar data items are close in the representation space
while pairs of dissimilar data items are far apart.

After the pre-training process, the learned data representation model is accessible to downstream tasks 
either directly in an unsupervised manner or by \emph{fine-tuning} using a few labels.
Unlike that in language models, fine-tuning is not always necessary if the downstream task is
similarity-based for which \system already provides a reliable metrics space. 
When fine-tuning is needed, \system provides an optional
\emph{pseudo labeling} step that 
extracts training signals with high confidence
from the learned representations useful for
further boosting the fine-tuning performance.
As we will illustrate, pseudo labeling together
with a customized fine-tuning architecture
significantly improves \system's performance on \di tasks
in low-resourced settings.

\begin{example}
We illustrate the \system pipeline in Figure \ref{fig:em}
using Entity Matching (EM), a typical application of
data integration. 
The upper part of the figure shows a standard EM pipeline
consisting of the blocking and matching sub-tasks.
Given two collections of entity entries (Tables A and B),
the goal of blocking is to select a candidate set of 
matching pairs from the cross product
to avoid the quadratic-size pairwise comparison.
Next, it trains a matching model 
on a labeled dataset sampled from the candidate set,
before applying the model to all the candidate pairs
for the final matching results.

\system starts with contrastive learning
to learn the similarity-aware data representations
of all entity entries from the two input tables.
Because its model already captures entity similarity,
we can solve the blocking task by
applying nearest neighbor search to identify
the top-$k$ most similar entity entries for all input entries
as the candidate set. 
For the matching task, practitioners can then
fine-tune the pre-trained representation model on
a labeled training set. For this step,
\system's pseudo labeling module allows automatic creation
of additional high-quality labels by extracting 
match/unmatch pairs with high confidence from the
learned representations. By doing so, \system significantly
reduces the label requirement for training the matching
model.

\begin{figure}[!t]
    \centering
    \includegraphics[width=0.46\textwidth]{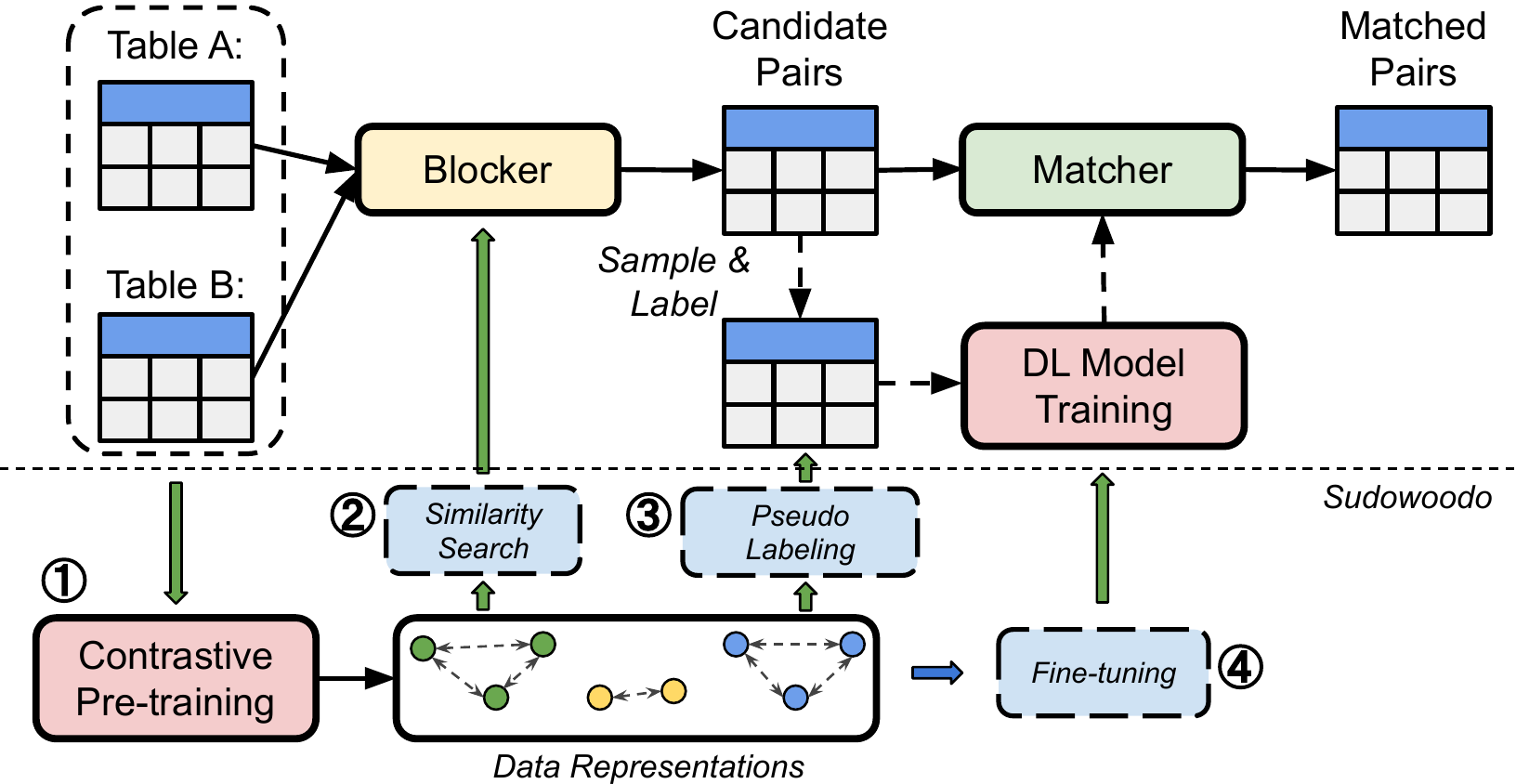}
    \caption{\small The \system pipeline for EM (blocking+matching).}
    \label{fig:em}
    \vspace{-5mm}
\end{figure}

\end{example}


\smallskip
\noindent
\textbf{Further optimizations. } 
Based on the standard contrastive learning framework, \system features several
optimizations to further boost the pre-training effectiveness.
First, the quality of learned representations heavily depends on the
generated pairs of data items that are semantic-preserving. 
\system obtain such pairs via data transformation such as synonym replacement or span deletion.
\rev{
To obtain more fine-grained transformations,
\system supports the \emph{cutoff} data augmentation (DA) technique~\cite{DBLP:journals/corr/abs-2009-13818}
that transforms input data items at the word-embedding level.
Combining cutoff with a package of \di data augmentation operators, \system obtains more diverse views for contrastive pre-training.}

Another challenge for applying contrastive learning to \di tasks is how to obtain more effective negative examples.
Negative examples allow the representation model to learn how to separate dissimilar
data items. The standard contrastive learning approach
uses uniformly sampled pairs as negative examples.
\rev{
This simple approach can be insufficient when random pairs
fail to provide strong enough negative examples because
they are easily separable.
To address this issue,
\system splits the unlabeled data items into clusters and 
samples negative pairs within each cluster.
By doing so, \system allows the representation model to learn ``harder'' by distinguishing
syntactically similar items.}

In summary, this paper makes the following contributions:
\begin{itemize}
\item We propose a generic matching problem definition 
that captures a wide variety of \di tasks including
(1) EM in data integration,
(2) Error correction in data cleaning, 
(3) Semantic type detection in data discovery, and more.
\item We propose \system, a contrastive representation learning framework for the supported \di tasks.
\system pre-trains a representation model from a large collection of unlabeled data items.
Practitioners can then apply the representations to 
downstream tasks either directly or by fine-tuning with few labels.
\item We introduce 3 optimizations to \system.
Specifically, we (1) augment the pre-training data using a novel cutoff operator,
(2) introduce ``harder'' negative samples via clustering, and (3) combine the standard contrastive loss with
a non-contrastive learning objective through redundancy regularization~\cite{DBLP:conf/icml/ZbontarJMLD21}.
\item We evaluate \system on real-world datasets for EM, data cleaning, and semantic type detection.
\system achieves a series of new SOTA results
under different label budgets, highlighting
its capability of holistically providing solutions for multiple \di tasks.
\end{itemize}

The rest of the paper is organized as follows. Section \ref{sec:overview} defines the problem
and introduces the overall architecture. 
We introduce the contrastive learning algorithms in Section \ref{sec:ssl}
and their optimizations in Section \ref{sec:optim}. Section \ref{sec:extend} extends \system
to data cleaning and column matching.
Section \ref{sec:experiment} summarizes the experiment results.
We discuss related work in Section \ref{sec:related} and conclude in Section \ref{sec:conclude}.

\section{Background and System Overview}\label{sec:overview}

In this section, we first define a general matching task for capturing
a broad set of \di tasks including Entity Matching (EM), data cleaning, and
column type detection. 
We then review the baseline method of pre-trained language models and 
fine-tuning for solving the task.
We also provide an overview of \system's architecture.

\subsection{Task definition}\label{subsec-task}

In \system, we consider \di tasks that can be formulated as a general 
problem of \emph{matching related data items}. 
These data items can be of diverse formats across different \di tasks. For example, 
as Figure \ref{fig:definition} illustrates,
these data items can be entity entries, 
table columns, cell values, etc.
The matching criteria can also be customizable 
binary relations w.r.t the specific \di applications,
such as: 
whether two entity entries refer to the same real-world entity for EM;
whether a row is a correct candidate correction
to an erroneous row for data cleaning; or 
whether two table columns have 
the same semantic type for data discovery.




It is commonly seen that within a \di task, the same matching problem is 
formulated in different ways to address different application needs. For example, 
in an EM task, the blocking stage requires quickly getting a small candidate set that contains filtered matching pairs
with high recall while the matching stage requires both high recall and precision.
To facilitate their different needs, in \system, we consider two flavors of the matching tasks: 
\emph{embedding} and \emph{pairwise matching}. 
The embedding task aims at generating a vector representation for each data item
so that related data items are close in the vector space. 
The generated vectors can then be indexed and queried to support fast similarity search~\cite{DBLP:conf/sigmod/Echihabi20}
or clustering~\cite{DBLP:conf/esws/SaeediPR18}. 
The goal of the pairwise matching task is to perform binary classification
for determining whether a pair of data items are related or not. We formally define these two tasks as follows:

\newcommand{\emb}{\mathsf{emb}}
\newcommand{\pmm}{\mathsf{pm}}

\begin{definition}[Embedding]
Given a collection $D$ of data items, a $d$-dimensional embedding model $M_{\emb}$ 
takes every data item $x \in D$ as input and outputs a real vector $M_{\emb}(x) \in \mathbb{R}^d$.
Given a similarity function $\mathsf{sim}$, e.g., cosine, for every pair of data items $(x, x')$, 
the value of $\mathsf{sim}(x, x')$ is large if and only if $(x, x')$ matches.
\end{definition}

For simplicity, we assume all output vectors are normalized, i.e. the $L$-2 norm $\Vert M_{\emb}(x) \Vert_2 = 1$ for all data item $x$.

\begin{definition}[Pairwise Matching]
Given a collection $D$ of data items, a pairwise matching model $M_{\pmm}$
takes a pair $(x, y)$ of data items as input and outputs $M_{\pmm}(x, y) \in \{0, 1\}$
where $M_{\pmm}(x, y) = 1$ if $(x, y)$ is a real match or $0$ otherwise.
\end{definition}

Note that we can also abuse the notation $M_{\pmm}(x, y)$ as a 2-d real vector
where each dimension indicates the predicted probability of non-match / match denoted as $M_{\pmm}(x, y)_0$ and $M_{\pmm}(x, y)_1$ respectively.

\smallskip
\noindent
\textbf{Entity Matching. }
With the help of \system, it is rather easy to build an effective EM pipeline.
The embedding model provides the similarity measure for the blocking stage of EM while
the matching stage can directly use the pairwise matching model that \system provides. 
For the rest of the paper, we will take EM as the main example when introducing the \system framework for clarity.
We will describe how to extend \system to the data cleaning and column type detection tasks later in Section \ref{sec:extend}. 


\subsection{Fine-tuning language models}\label{subsec-lm}

The goal of \system is to (1) train high-quality embedding models without labels
and (2) train the pairwise matching models using no or just a few labels.
Pre-trained language models (LMs), such as BERT, provide a good baseline solution 
which has recently achieved promising results for
EM~\cite{ditto2021,DBLP:journals/pvldb/PeetersB21,DBLP:journals/pvldb/ThirumuruganathanLTOGPFD21,li2021deep}.
These models typically consist of 12 or more Transformer layers~\cite{DBLP:conf/nips/VaswaniSPUJGKP17}
pre-trained on large unlabeled text corpora such as Wikipedia.
In the process of pre-training, the LMs are trained in a self-supervised manner 
to perform auxiliary tasks such as missing token and next-sentence prediction.
In this way, LMs effectively capture lexical or semantic meanings of 
the input text sequences. 
Formally, a LM $M$ encodes an input token sequence $x$ into a vector representation $M(x)$.

To apply pre-trained LMs in \system, we first need to serialize the input data items
into sequences of tokens. For EM, we follow the serialization method 
in Ditto~\cite{ditto2021} which converts an entity entry into the format:

\begin{small}
\begin{center}
[COL] Title [VAL] instant immers \dots \ [COL] Price [VAL] 36.11
\end{center}
\end{small}

\noindent
where [COL] and [VAL] are special tokens indicating the start of 
an attribute or value. We denote by $\mathsf{serialize}(x)$ the serialization
of a data item $x$. We omit ``$\mathsf{serialize}$'' when the context is clear.
After that, we can obtain a baseline embedding model
that first serializes an input data item and encodes it 
using the pre-trained LM.

Next, one can obtain a baseline pairwise matching model via \emph{fine-tuning}
the pre-trained model on a downstream task-specific training set. Pre-trained LMs
like BERT support sequence pair classification by concatenating the input pair 
$(x, y)$ into a single sequence as

\begin{center}
[CLS] $\mathsf{serialize}(x)$ [SEP] $\mathsf{serialize}(y)$ [SEP]
\end{center}

\noindent
where the [SEP]'s are separator tokens and [CLS] is a special token to
obtain the sequence pair representation. Fine-tuning typically consists of the following
steps:
\begin{itemize}
\item Add task-specific layers after the last Transformer layer. 
In our case, the task-specific layers consist of a linear fully connected layer and a softmax layer for binary classification.
\item Initialize the modified network with the pre-trained weights.
\item Train the modified network until convergence.
\end{itemize}

Since the LM is pre-trained to gain semantic understanding capability, 
it is expected that the fine-tuning approach requires fewer labels
than training the model entirely from scratch.

\vspace{-1mm}
\subsection{Overview of \system}\label{subsec-system}

The pre-trained LM approach works reasonably well, but it is sub-optimal in the \di tasks supported by \system.
The main reason is that the pre-training tasks do not train the LM to explicitly capture the similarity between data items.
For the embedding model (i.e., blocker),
without further fine-tuning~\cite{DBLP:conf/emnlp/ReimersG19,DBLP:journals/pvldb/DengSL0020,DBLP:journals/pvldb/ThirumuruganathanLTOGPFD21}, 
the LM representations for two closely related data items
may not be sufficiently close to support effective similarity search.
The fine-tuning process of concatenating input data items
also does not model the ``differences'' between data items
potentially leading to lower matching accuracy or a higher label requirement.

To address this challenge, \system utilizes the
\emph{contrastive learning}~\cite{DBLP:conf/icml/ChenK0H20} technique.
Unlike LM pre-training, contrastive learning explicitly captures relatedness between data items.
The pre-training objective is designed to enforce vector representations of 
similar data items to be close and different items to be far apart.
We will introduce the contrastive learning algorithm in Section \ref{sec:ssl}
and its optimizations in Section \ref{sec:optim}.

The lower part of Figure \ref{fig:em} illustrates the overall framework of \system in the EM application.
\system starts with a pre-training phase \textcircled{1} with contrastive learning. 
During pre-training, \system continuously updates a pre-trained LM using 
positive/negative pairs of data items created by data transformation and 
negative sampling methods. This process produces an embedding model more suitable to be used
 in step \textcircled{2}. 
For blocking, we apply the embedding model to vectorize each data item and
use high-dimensional similarity search technique to filter out the $k$ nearest
items for every input as the candidate set.

To further reduce the label requirement of fine-tuning, in step \textcircled{3},
\system applies pseudo labeling techniques (Section \ref{sec:pseudo}) 
to automatically generate probabilistic 
training examples. These examples are combined with the manually labeled ones
to form the training set for the fine-tuning step \textcircled{4}. 
For fine-tuning, \system features a new task-specific layer 
that captures the difference between data representations beyond simple concatenation.

\smallskip
\noindent
\textbf{Discussion about scalability}. The design of the
\system pipeline is optimized towards delivering 
high-quality matching results. In practice, 
one might train and deploy the models in different manners 
to achieve a good balance between quality and scalability,
which is beyond the scope of this work.
However, we notice that by utilizing \system's learned representations, 
the blocking stage tends to produce a significantly smaller candidate set with high recall that leads to
acceleration of the matching stage.



\section{Contrastive Learning}\label{sec:ssl}

We overview contrastive learning in Section
\ref{sec:simclr} and explain why directly
applying the SimCLR algorithm
is not sufficient for \system's 
matching problem settings.
We propose two novel techniques to address
the issue:
(1) a customized fine-tuning approach 
and (2) pseudo-labeling in 
Section \ref{sec:finetune}
and \ref{sec:pseudo} respectively.





\vspace{-1mm}
\subsection{Representation learning with SimCLR} \label{sec:simclr}

\begin{figure}[!t]
    \centering
    \includegraphics[width=0.35\textwidth]{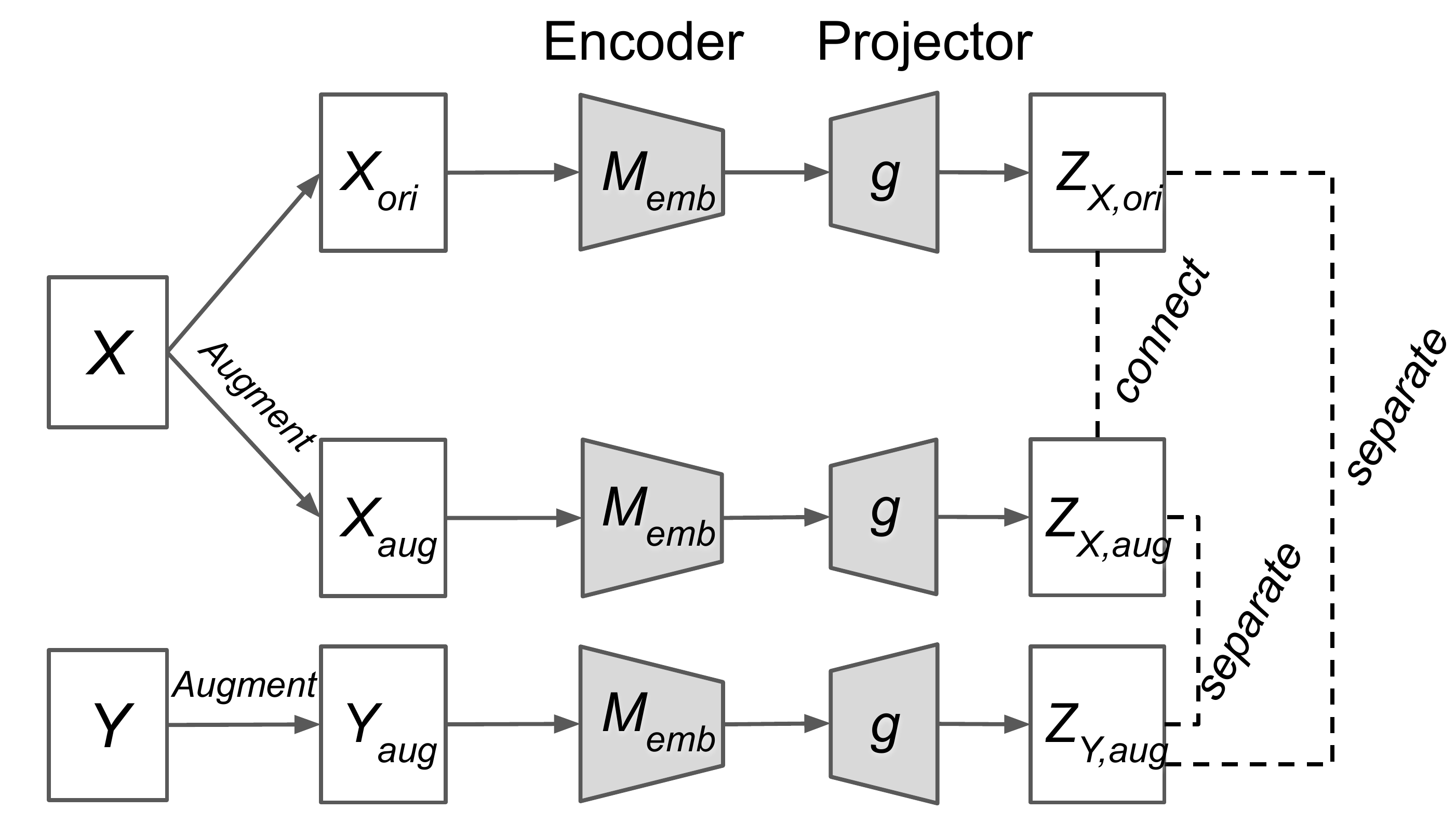}
    \caption{The SimCLR algorithm pre-trains representations (e.g., $z_{x,ori}$, $z_{x,aug}$, and $z_{y,aug}$) by connecting 
    similar data items (e.g., $x_{\mathsf{ori}}$ and $x_{\mathsf{aug}}$) 
    and separating distinct items (e.g., $x_{\mathsf{aug}}$ and $y_{\mathsf{aug}}$).}
    \label{fig:simclr}
    \vspace{-3mm}
\end{figure}

Contrastive learning is a self-supervision approach that learns data representations where similar data items are close while 
different ones are far apart. 
In \system, we choose SimCLR~\cite{DBLP:conf/icml/ChenK0H20}, 
which has shown effectiveness in learning visual representations and NLP tasks~\cite{DBLP:conf/emnlp/GaoYC21}, 
as the base algorithm.
At a high level, SimCLR pre-trains a representation model
by simultaneously (1) minimizing distance between pairs of same/similar data items
and (2) maximizing distance between pairs of distinct items
as Figure \ref{fig:simclr} illustrates. 

To achieve this goal without labels, for objective (1), 
we utilize data augmentation operators
to generate variants (i.e., multiple views) of the same data item.
These operators are task-specific data transformations that preserve the semantics of 
the input data item. For example, in EM, such transformations can be permuting 
the attribute order or replacing a token with its synonym.

Minimizing objective (1) only can result in a trivial solution where
all representations collapse into a single vector.
SimCLR leverages \emph{negative sampling} to address this issue in objective (2).
The default negative sampling method uses 
uniformly sampled pairs as negative pairs. 
This is based on the observation that two randomly sampled 
items are likely to be distinct in a large corpus. 

We outline the pseudo code of SimCLR in Algorithm \ref{alg:simclr}.
The algorithm consists of four major steps:

\begin{itemize}
    \item Data augmentation (DA, Line 7). This module contains a set of DA operators that 
    generate semantically equivalent distorted views ($B_{\mathsf{ori}}$ and $B_{\mathsf{aug}}$) of the same data item. 
    \item The embedding model $M_{\emb}(\cdot)$ is 
    an encoder for extracting representation vectors from data items (Line 1). 
    In \system, we design the encoder to be a Transformer-based pre-trained LM such as BERT and its variants. 
    We note that SimCLR uses the same encoder for both the original and augmented data items.
    \item A small projection head $g(\cdot)$ (Line 2).
    The projector maps representations to the space where the contrastive loss is applied. 
    We use a linear layer as the projection head while vision tasks use a more complicated design~\cite{DBLP:conf/icml/ChenK0H20}.
    Both the encoder and projector are trainable and updated during the learning process (Line 10).
    \item A contrastive loss function (Line 9). See details below.
\end{itemize}


The contrastive loss function works as follows. Let $N$ be the batch size and $Z = \{z_i\}_{1 \leq i \leq 2N}$
be the concatenation of the two encoded views $Z_{\mathsf{ori}}$ and $Z_{\mathsf{aug}}$. 
Here $z_i$ is the $i$-th element of $Z_{\mathsf{ori}}$ for $i \leq N$ and the $(i-N)$-th element of $Z_{\mathsf{aug}}$ for $i > N$.
We first define a single-pair loss $\ell(i, j)$ for an element pair $(z_i, z_j)$ to be
\vspace{-2mm}
\begin{equation}\label{equ:infonce1}
\vspace{-0mm}
\ell(i, j)=-\log \frac{\exp \left(\operatorname{sim}\left(\boldsymbol{z}_{i}, \boldsymbol{z}_{j}\right) / \tau\right)}{\sum_{k=1}^{2 N} \mathds{1}_{[k \neq i]} \exp \left(\operatorname{sim}\left(\boldsymbol{z}_{i}, \boldsymbol{z}_{k}\right) / \tau\right)}
\end{equation}
where \emph{$\operatorname{sim}$} is a similarity function (e.g., cosine) and $\tau$ is a temperature hyper-parameter in the range $(0, 1]$.
The single-pair loss is applied to every pair of $(z_i, z_j)$ that are the same views, such as $(z_1, z_{N+1})$.
By minimizing this loss, we maximize the similarity term $\operatorname{sim}\left(\boldsymbol{z}_{i}, \boldsymbol{z}_{j}\right)$ in the numerator
and minimize $z_i$'s similarity values with other elements 
in the denominator.
We can then obtain the contrastive loss $\mathcal{L}$ by averaging all matching pairs:
\begin{equation}\label{equ:infonce2}
\mathcal{L}_{\mathsf{contrast}} =\frac{1}{2 N} \sum_{k=1}^{N}[\ell(k, k+N)+\ell(k+N, k)].
\vspace{-1mm}
\end{equation}

\SetKwInOut{Parameter}{Variables}
\begin{algorithm}[!t]
\small
	\KwIn{ A dataset $D$ of serialized data items}
	\Parameter{ Number of training epochs $\mathsf{n\_epoch}$; \\
	       Data augmentation operator $\mathsf{op}$; Learning rate $\eta$ }
	\KwOut{ An embedding model $M_{\emb}$
	        }
	\tcc{encoder and projector}
	Initialize $M_{\emb}$ using a pre-trained LM\;
	Initialize $g$ randomly\;
	$M \leftarrow g \circ M_{\emb}$ \tcp*{appending $g$ after $M_{\emb}$}
	\For{ $\mathsf{ep} = 1$ to $\mathsf{n\_epoch}$}{
	    Randomly split $D$ into mini-batches $\{B_1, \dots B_n\}$\;
	    \For{$B \in \{B_1, \dots B_n\}$} {
	        \tcc{augment and encode every item}
	        $B_{\mathsf{ori}}, B_{\mathsf{aug}} \leftarrow \mathsf{augment}(B, \mathsf{op})$\;
	        $Z_{\mathsf{ori}}, Z_{\mathsf{aug}}  \leftarrow M(B_{\mathsf{ori}}), M(B_{\mathsf{aug}})$\;
	        \tcc{Equation (\ref{equ:infonce1}) and (\ref{equ:infonce2})}
	        $\mathcal{L} \leftarrow \mathcal{L}_{\mathsf{contrast}}(Z_{\mathsf{ori}}, Z_{\mathsf{aug}})$\;
	        \tcc{Back-prop to update $M_\emb$ and $g$}
	        $M \leftarrow \textsf{back-propagate}(M, \eta, \partial \mathcal{L} / \partial M) $\;
	    }
	}
	\Return $M_{\emb}$, discard $g$\;
	\caption{$\textsf{SimCLR pre-training}$}
	\label{alg:simclr}
\end{algorithm}




However, directly applying contrastive learning algorithms 
as described above to the pairwise
matching problem settings of \di discussed in this paper might result in sub-optimal performance. 
The main reason is that after pre-training, 
the model captures representations of single data items,
while the downstream tasks of \di need to handle pairs of data items 
as input. Without special treatment, 
the default fine-tuning option 
cannot make full use of the rich semantics captured during pre-training.
Next, we introduce two novel techniques 
to address this challenge.

\subsection{Fine-tuning in \system}\label{sec:finetune}

After pre-training, the embedding model captures similarity among data items. 
For pairwise matching, the default option of LM fine-tuning is to first concatenate
the input pair of serialized data items and then encode the concatenated sequence 
using $M_{\emb}$. 
This is not ideal because $M_{\emb}$ is pre-trained to encode single items
instead of concatenated pairs. 
On the other hand, encoding the two data items separately
can miss cross entity information (e.g., the difference between two product ID spans),
which can be captured by the LM's self-attention mechanism applied 
to the concatenated pairs.

\begin{figure}[!t]
    \vspace{-1mm}
    \centering
    \includegraphics[width=0.48\textwidth]{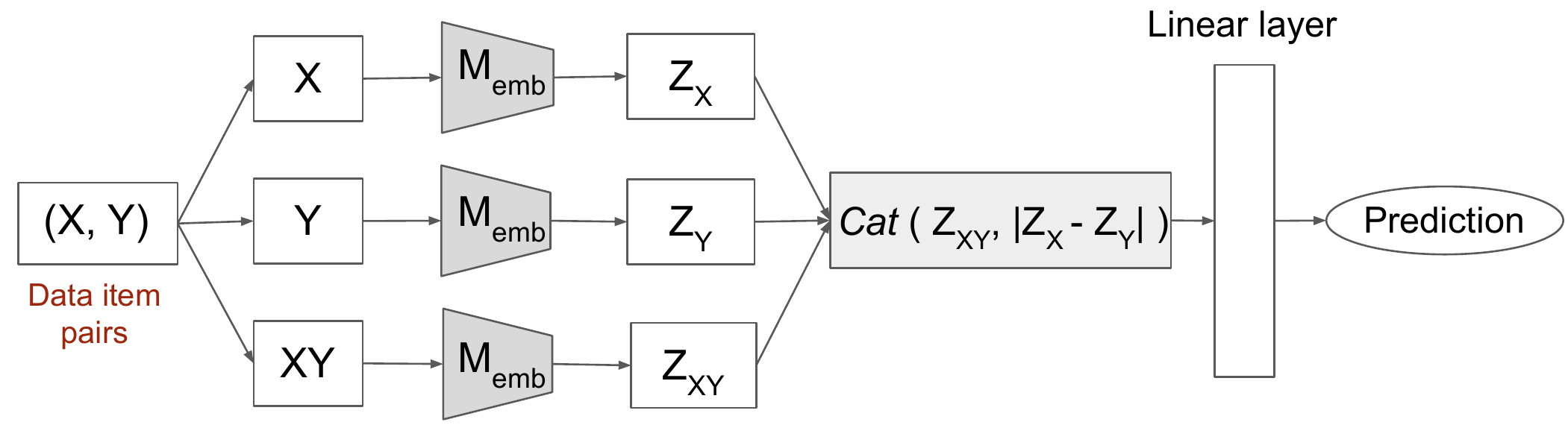}
    \caption{Fine-tuning for pairwise matching in \system. 
    }
    \label{fig:finetune}
    \vspace{-4mm}
\end{figure}

In \system, we extend the default fine-tuning with an extra step
for capturing similarity/difference between the input pairs.
We illustrate the fine-tuning model architecture in Figure \ref{fig:finetune}.
Specifically, given a pair $(x, y)$ of serialized data items,
we applied the pre-trained representation model $M_\emb$
on $x$, $y$ individually and on the concatenation $xy = \mathsf{concat}(x, y)$.
Let $Z_x$, $Z_y$ and $Z_{xy}$ be the $d$-dimensional representations
$M_\emb(x)$, $M_\emb(y)$, and $M_\emb(xy)$.
The task-specific layers of \system consist of a linear layer 
$\text{Linear}_{\mathsf{diff}}$
of input dimension $2d$ and a softmax function that predicts whether the pair matches or not (0/1). 
The pairwise matching model $M_{\pmm}$ is defined as
\begin{equation}\label{equ:finetine}
M_{\pmm}(x, y) = \mathsf{softmax}\left(\text{Linear}_{\mathsf{diff}}( Z_{xy} \oplus |Z_x - Z_y| ) \right) 
\end{equation}
where $\oplus$ denotes vector concatenation. Computing the vector subtraction and absolute values is done element-wise.


\subsection{Pseudo labeling}\label{sec:pseudo}


In addition to fine-tuning, \system also utilizes a pseudo labeling step
to extract similarity-based knowledge from the learned data item to further boost the performance of pairwise matching problems.
This is realized by automatically generating probabilistic labels to augment the manually labeled set.
It works for pairwise matching tasks that are similarly-based,
for which the embedding model $M_\emb$ provides a reliable resource for
testing whether a pair of data items are similar or not.

At a high level, for every unlabeled pair $(x, y)$ in the candidate set,
\system measures the confidence of whether $x$ and $y$ matches
by the cosine similarity between their representations $M_\emb(x)$ and $M_\emb(y)$.
We assign the pair the positive label if the cosine similarity
$\cos(M_\emb(x), M_\emb(y))$ is above a positive threshold $\theta^+$ and 
the negative label if their similarity is below a negative threshold $\theta^-$.
The choices for hyper-parameters $\theta^+$ and $\theta^-$ are crucial as they determine
the quality and size of the pseudo labels. For example, a more relaxed combination of 
$(\theta^+, \theta^-)$ can result in a training set of larger size but more noise in
the labels.

Yet another important factor to consider is balancing the ratio between
positive and negative labels. \di tasks such as EM or data cleaning naturally
have much more negative labels (i.e., non-match pairs / clean cells) than positive ones.
A training set with a high ratio of negative pairs can result in a matching model that is
heavily biased towards predicting negative during fine-tuning. 

In \system, we take a semi-automatic approach to tune $(\theta^+, \theta^-)$ as
hyper-parameters. The user first needs to fix a positive ratio $\rho$ from a small set
$\{5\%, 10\%, \dots \}$. This ratio can also be estimated using a few uniformly 
sampled labels.
Let $C$ be the candidate set and $C^+$ / $C^-$ be the subset of $C$ above $\theta^+$ 
or below $\theta^-$ respectively. By fixing $\rho$, we would like to keep 
${|C^+|} / {(|C^+| + |C^-|)} = \rho$ and only search for $\theta^+$
(or $\theta^-$).
We then take a hill-climbing heuristics~\cite{DBLP:conf/kdd/AkibaSYOK19}
to find a locally optimal $\theta^+$ using a fixed number of fine-tuning trials.

\section{Pre-training optimizations}\label{sec:optim}

By the design of the contrastive learning framework, 
the quality of the representations depends on the quality of 
(1) data augmentation (DA) operators for generating positive pairs and 
(2) negative sampling for generating distinctive pairs.
For (1), we propose a task-specific DA method with cutoff~\cite{DBLP:journals/corr/abs-2009-13818}
which perturbs the input token embeddings of the LM directly to generate augmentations with more diversity.
For (2), we propose a clustering-based sampling technique that samples negative pairs from the same cluster.
We expect the negative samples to become ``harder'' to distinguish thus the representation model is forced to
learn more meaningful features.

In \system, we also improve the pre-trained model by combining contrastive learning
with Barlow Twins~\cite{DBLP:conf/icml/ZbontarJMLD21}, a recently proposed self-supervised learning technique
based on feature redundancy regularization. Without relying on negative sampling, redundancy regularization
enforces the learned features in the representation space to be independent of each other
so that it avoids having multiple features capturing the same entity information. 

\subsection{Data augmentation (DA) with cutoff}\label{sec:cutoff}

The original goal of DA is to generate additional training
examples from existing ones. DA operators are typically data transformations that
preserve the semantic meaning of an input data item. In computer vision, these transformations
can be rotation, flipping, or cropping of an image. DA has also been
applied to learning-based \di tasks (see \cite{DBLP:journals/pvldb/00010MT21} for a survey).
In contrastive learning, DA has a slightly different goal of generating pairs
of data items that are similar to each other. It is, therefore, less of an issue of getting corrupted labels
(e.g., identical data items becoming distinct) as long as the transformation preserves a certain level of data similarity.

Based on this intuition, we design the DA operators in \system by combining a base set of task-specific operators
with additional perturbation using the cutoff operator~\cite{DBLP:journals/corr/abs-2009-13818}.
Following previous works~\cite{ditto2021,DBLP:conf/sigmod/Miao0021}, we support 
the set of DA transformation operators listed in Table \ref{tab:defaultda} for EM.
The supported operators include token-based, span-based, and attribute-based transformations.
In this work, we apply a single base DA operator at a time.
We note that one might also apply a learning-based
DA such as Rotom~\cite{DBLP:conf/sigmod/Miao0021} to
automatically select and combine multiple 
DA operators, which we leave as future work.

\setlength{\tabcolsep}{5.5pt}
\begin{table}[!t]
    \caption{DA operators in \system for EM. }\label{tab:defaultda}
{\small
\begin{tabular}{ll}\toprule
\textbf{Operators}     & \textbf{Details}                                                       \\ \midrule
token\_del    & Sample and delete a token                                     \\
token\_repl   & Sample a token and replace it with a synonym                  \\
token\_swap   & Sample two tokens and swap them                               \\
token\_insert & Sample a token and insert to its right a synonym \\
span\_del     & Sample and delete a span of tokens                            \\
span\_shuffle & Sample a span of tokens and shuffle their order               \\
col\_shuffle  & Choose two attributes and swap their order            \\
col\_del      & Choose an entity attribute and drop it entirely                \\ \bottomrule
\end{tabular}}
\vspace{-1mm}
\end{table}

\begin{figure}[!t]
    \centering
    \includegraphics[width=0.48\textwidth]{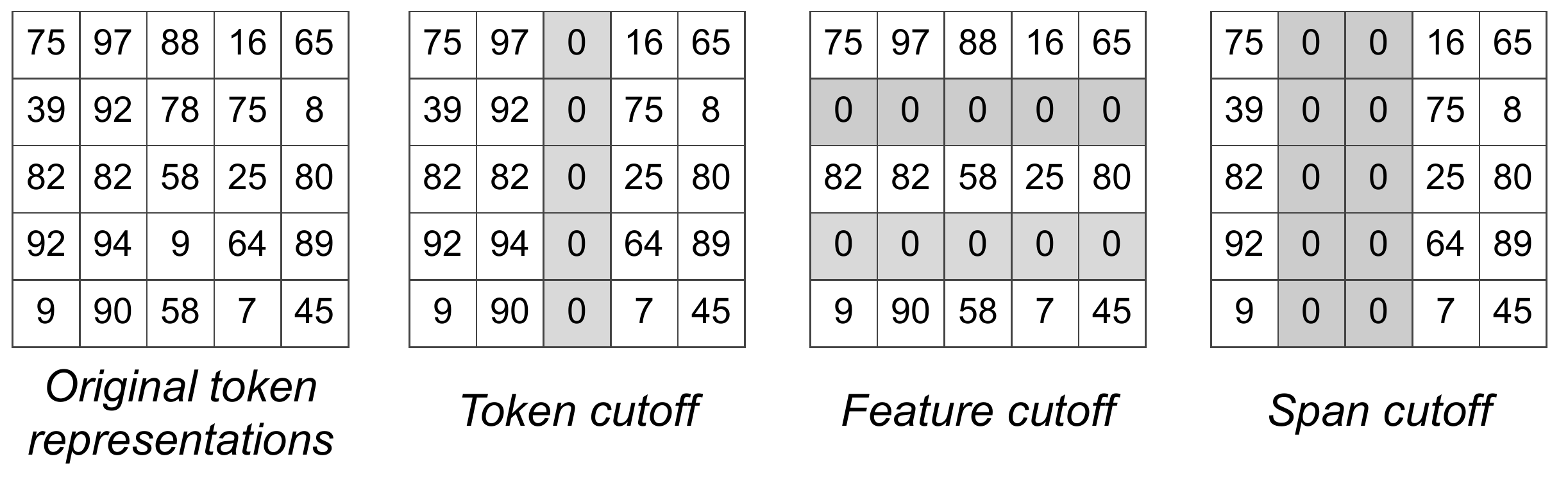}
    \caption{\small Three cutoff DA operators. The cutoff operators
    modify rows/columns of the input token embedding matrices directly.}
    \label{fig:cutoff}
	\vspace{-4mm}
\end{figure}

\system then combines the base transformations with a cutoff operator. 
The cutoff operators directly modify the input token embeddings to the representation model. 
We consider 3 cutoff operators: token-cutoff, feature-cutoff, and span-cutoff. 
As Figure \ref{fig:cutoff} illustrates, given an input serialized data item,
after converted to a sequence of token embeddings (each column is the embedding of an input token),
these 3 operators uniformly sample a token index, a set of feature indices, or a span of token indices 
and set the corresponding dimensions in the embedding matrix to zeros.

\system performs the cutoff operators batch-wise, i.e., the same token/feature/span-cutoff is applied
to all data items in the same batch. This allows the cutoff to gain additional effects that
at each training step, the Transformer-based encoder uses only partial information to make the matching prediction.
It prevents the encoder to ``overfit'' certain features or parts of the input content.
A similar design idea is used in the popular dropout mechanism~\cite{DBLP:journals/jmlr/SrivastavaHKSS14} 
for training deep neural networks.

\subsection{Clustering-based negative sampling}\label{sec:clustering}

As mentioned in Section \ref{sec:ssl}, the default sampling method uses pairs of uniformly sampled data items
as negative examples. This method can be insufficient because it can be quite easy to differentiate
such pairs, e.g., by checking if the data items contain overlapping tokens, without capturing the more important
feature such as product ID's. 

To address this challenge, \system generates more effective negative examples by sampling pairs 
of data items that are \emph{lexically} similar. By doing so, we can train the representation model
``harder'' and force it to capture the semantic meaning of data items beyond lexicons. 
We use the standard TF-IDF (Term frequency - Inverse document frequency) method to
obtain a sparse vector representation for each data item. 
We then use the cosine similarity of the sparse vectors to measure the lexical similarity between each pair.

The process for generating similar pairs needs to be efficient since the input unlabeled corpus can be quite large
and we need to repeatedly sample training batches for every training epoch.
To this end, we choose the k-means algorithm which has running time linear to the dataset size $|D|$
and the number of clusters $k$. We add additional shuffling steps among and within the clusters
to generate batches with more diversity. The negative sampling procedure is outlined in Algorithm \ref{alg:negative}
(replacing Line 5 of Algorithm \ref{alg:simclr}).



\begin{algorithm}[!t]
\small
	\KwIn{ A dataset $D$ of serialized data items }
	\Parameter{ Number of clusters $k$; Batch size $N$ }
	\KwOut{ A set of mini-batches $\{B_1, B_2, \dots B_{\lceil|D| / N \rceil}\}$ }
	\tcc{Generate TF-IDF features}
	$F \leftarrow \text{TF-IDF-Featurize}(D)$\;
	\tcc{Clustering and shuffling. Cache the results for future epochs}
	$\mathsf{Clusters}\leftarrow \textsf{k-means}(D, F, k)$\;
	$\mathsf{Clusters}\leftarrow \textsf{shuffle}(\mathsf{Clusters})$\;
	$\mathsf{Batches} \leftarrow \emptyset$\;
	
	\For{$C \in \mathsf{Clusters}$}{
	    $C \leftarrow \textsf{shuffle}(C)$\;
	    \For{$x \in C$}{
	        \tcc{Add to the last batch}
	        $B_{\mathsf{Last}} \leftarrow B_{\mathsf{Last}} \cup \{x\}$\;
	        \If{$|B_{\mathsf{Last}}| = N$}{
	            $\mathsf{Batches} \leftarrow \mathsf{Batches} \cup \{B_{\mathsf{Last}}\}$\;
	            $B_{\mathsf{Last}} \leftarrow \emptyset$\;
	        }
	    }
	}
	\Return $\mathsf{shuffle(Batches)}$\;
	\vspace{-1mm}
	\caption{$\textsf{Clustering-based negative sampling}$}
	\label{alg:negative}
	
\end{algorithm}


\subsection{Redundancy regularization}\label{sec:barlow_twins}
 
Redundancy regularization is a recently proposed technique for self-supervised representation learning.
\rev{Unlike contrastive learning, redundancy regularization does not rely on negative sampling to avoid the
potential issue of collapsing into a trivial solution. 
To achieve the same effect,
it aims at learning data representations where features are \emph{orthogonal} to each other. 
By doing so, the learned features can capture
different aspects of the data items to provide 
high-quality representations for measuring similarity.
Representations are also unlikely to collapse into trivial
solutions if they are different in any aspects captured by the orthogonal features.}

\begin{figure}[!t]
    \centering
    \includegraphics[width=0.48\textwidth]{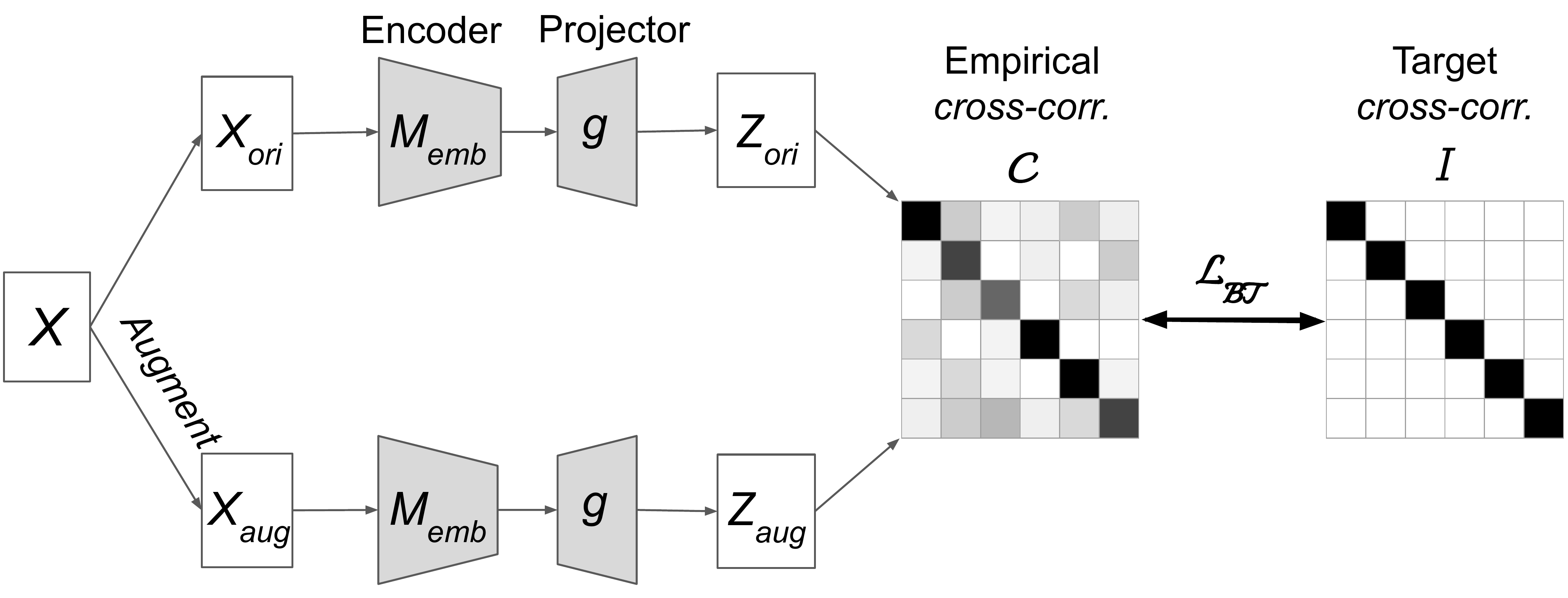}
    \caption{Redundancy regularization with Barlow Twins (BT). 
    The learning goal of BT is to learn orthogonal representation features 
    such that the cross-correlation matrix is close to an identity matrix.}
    \label{fig:barlow_twins}
    \vspace{-3mm}
\end{figure}

Barlow Twins (BT) is the representative method that leverages redundancy regularization. 
The pre-training algorithm of BT has an identical structure with the SimCLR algorithm as Figure \ref{fig:barlow_twins} illustrates.
Similar to SimCLR, the BT algorithm first applies DA operators to generate two different views 
$X_{\mathsf{ori}}$ and $X_{\mathsf{aug}}$ of the same batch $X$ of data items. BT applies the encoder $M_\emb$
and projector $g$ to obtain vector representations $Z_{\mathsf{ori}}$ and $Z_{\mathsf{aug}}$
\rev{(i.e., Line 8 of Algorithm \ref{alg:simclr})}.

Next, the BT algorithm computes the \emph{cross-correlation matrix} between $Z_{\mathsf{ori}}$ and $Z_{\mathsf{aug}}$.
Intuitively, the cross-correlation matrix captures the cosine similarity between all pairs of \emph{features}
instead of data items as in SimCLR. Here, on a batch of data representations of size $N$, 
a feature is represented by a $N$-dimensional vector consisting of the feature's values within the batch.
Formally, for a $d$-dimensional representation space, the cross-correlation matrix is a $d$-by-$d$ matrix
where each element is defined as
\vspace{-2mm}
\begin{equation}\label{equ:cross_correlation}
\mathcal{C}_{i j} := \dfrac{\sum_{b=1}^N z_{b, i}^{\mathsf{ori}} z_{b, j}^{\mathsf{aug}}}{\sqrt{\sum_{b=1}^N \left(z_{b, i}^{\mathsf{ori}}\right)^{2}} \cdot \sqrt{\sum_{b=1}^N\left(z_{b, j}^{\mathsf{aug}}\right)^{2}}}
\vspace{-1mm}
\end{equation}
for dimension $i$ and $j$ in the representation space. Here an element $z_{b, i}^{\mathsf{ori}}$ or $z_{b, i}^{\mathsf{aug}}$ refers to
the $i$-th dimension value of the $b$-th row in the batch $Z_{\mathsf{ori}}$ and $Z_{\mathsf{aug}}$ respectively.
By this definition, the element $\mathcal{C}_{i j}$ captures the cosine similarity between feature $i$ and $j$.

In order for the representations to be effective, we would like the same feature to be strongly correlated to itself
in the two different views so the diagonal values $\mathcal{C}_{i i}$ should be as close to 1 as possible. 
On the other hand, different features should be orthogonal to each other 
thus the off-diagonal values $\mathcal{C}_{i j}$ for $i \neq j$
should be close to 0. To sum up, to learn effective representations, the cross-correlation matrix should be close to the $d$-dimensional
identity matrix $I$. The loss function of BT is then defined as:
\begin{equation}
\vspace{1mm}
    \mathcal{L}_{\mathsf{BT}} := \underbrace{\sum_{i=1}^d\left(1-\mathcal{C}_{i i}\right)^{2}}_{\text {invariance term }}\quad + \underbrace{\lambda\sum_{i=1}^d \sum_{j \neq i} \mathcal{C}_{i j}^{2}}_{\text {redundancy reduction term }} 
    \vspace{-1mm}
\end{equation}
where $\lambda$ is a hyper-parameter balancing the weights of terms.


\rev{
To integrate BT into the pre-training process, we can combine
the contrastive loss (Equation \ref{equ:infonce2}) with
$\mathcal{L}_{\mathsf{BT}}$ linearly.
Formally, the combined loss function is
\begin{equation}\label{eq-alpha}
\vspace{-2mm}
    \mathcal{L}_{\text{\system}} :=  (1 - \alpha) \mathcal{L}_{\mathsf{Contrast}} + \alpha \mathcal{L}_{\mathsf{BT}} 
\end{equation}
where $\alpha$ is the hyper-parameter controlling the weights of two loss functions.
For the pre-training algorithm,
we only need to replace Line 9 of Algorithm
\ref{alg:simclr} with Equation 
\ref{eq-alpha} to compute the combined loss
$\mathcal{L}_{\text{\system}}$.
}

\section{Extension to More Tasks}\label{sec:extend}

We have introduced how to support different sub-tasks in Entity Matching with \system
in the previous sections.
In this section, we generalize \system to more \di tasks, namely
data cleaning and column type detection. 



\subsection{Data cleaning}\label{sec:cleaning}

Data cleaning refers to the process of detecting and correcting 
corrupt, inconsistent, or missing data records from dirty data sources
such as spreadsheets or relational tables. It is an important data preparation task
for improving data quality for 
downstream applications~\cite{DBLP:conf/sigmod/ChuIKW16,DBLP:conf/icde/LiRBZCZ21}.
A data cleaning pipeline typically consists of two stages: error detection (ED)~\cite{DBLP:conf/sigmod/MahdaviAFMOS019,DBLP:conf/sigmod/HeidariMIR19}
and error correction (EC)~\cite{DBLP:journals/pvldb/MahdaviA20}. 
While many existing works focus on optimizing only one stage at a time,
\system provides a holistic solution for generating data corrections
directly from the potentially contaminated data.

We consider the following problem setting. Given an input dirty relational table 
$T$ of $n$ rows with $m$ attributes, for each row $r \in T$ and 
each cell value $r_i \in r$, generate the correction 
$r_i'$ for $r_i$ (if dirty) given $(r, T)$ as the context.

As Section \ref{sec:overview} mentions, to apply \system to the data cleaning problem,
we first need a module that generates \emph{candidate corrections} 
for each record. \system follows the setting of 
Baran~\cite{DBLP:journals/pvldb/MahdaviA20} which uses multiple
external EC tools to generate the candidate corrections.
These tools cover a wide variety of error types including
missing value, typo, formatting issue, and violated attribute dependency (see Table \ref{tab:cleaning_dataset}).
Formally, for each cell value $r_i$, the EC tools generate a candidate
correction set $\mathsf{cand}(r_i)$. \revicde{We note that Baran is a feature-based active learning method while \system is a deep learning approach based on contrastive learning and language models. }

To apply \system in the data cleaning tasks, we first pre-train
a representation model $M_\emb$ with all cells and their candidate corrections as input.
We consider two serialization schemes: contextual and context-free as in \cite{DBLP:conf/sigmod/Miao0021}.
The context-free scheme serializes a cell $r_i$ as ``[COL] $\mathsf{attr}_i$ [VAL] $r_i$''
where $\mathsf{attr}_i$ is the name of the $i$-th attribute. The contextual scheme serializes an entire row $r$
into 

\begin{small}
\begin{center}
[COL] $\mathsf{attr}_1$ [VAL] $r_1$ $\dots$ [COL] $\mathsf{attr}_m$ [VAL] $r_m$
\end{center}
\end{small}
and replaces $r_i$ with a corrected value in $\mathsf{cand}(r_i)$ when encoding a candidate correction.
Similar to EM, all optimizations (data augmentation, negative sampling, and redundancy regularization) for pre-training
introduced in Section~\ref{sec:optim} can be applied to this step.

A ``blocking'' phase serves as an optional step for further refining the candidate sets. This is usually not needed
if the candidate sets are sufficiently small, but we note that there are cases where the refinement step is necessary.
For example, the correction generator may choose to replace a misspelled token with words from a large vocabulary or 
may fill a missing value using a large domain (e.g., all possible city names). 

Next, the domain experts label a few pairs of cell values with candidate corrections, i.e., $(r_i, r_i^\mathsf{c})$
where $r_i^\mathsf{c}$ is a correction from the set $\mathsf{cand}(r_i)$. The label is binary (1 or 0) indicating
whether the correction is right or not. For data labeling, we do not apply the pseudo labeling step 
since the task is not similarity-based. \system then fine-tunes the embedding model $M_\emb$ into
the pairwise matching model $M_\pmm$ using the same method introduced in Section~\ref{sec:finetune}.

To finalize the correction results, for each cell $r_i$, we find the correction $r_i^\mathsf{c}$ that maximizes the 1-probability
as the model $M_\pmm$ predicts, i.e., $r_i' = \argmax_{r_i^\mathsf{c} \in \mathsf{cand}(r_i)} M_\pmm(r_i, r_i^\mathsf{c})_1 $.
The cell is considered clean if $r_i = r_i'$ or $M_\pmm(r_i, r_i') = 0$. Otherwise, we output $r_i'$ as the correction for $r_i$.

\vspace{-2mm}

\subsection{\revicde{Semantic} type detection}\label{sec:column}

\revicde{Given a table with several columns, the semantic type detection task considers assigning semantic types, such as “city”, “age”, and “population”, to each column ~\cite{DBLP:journals/pvldb/CafarellaHWWZ08,DBLP:journals/pvldb/LimayeSC10,DBLP:conf/kdd/HulsebosHBZSKDH19,DBLP:conf/wsdm/ZhangWSDFP20}. Existing learning-based methods considers the multi-class classification setting where the classifier assigns a single semantic type to each column \cite{DBLP:conf/kdd/HulsebosHBZSKDH19, DBLP:journals/pvldb/ZhangSLHDT20}, and this formation is called as column type detection.}
It suffers from the challenge that domain experts need to conduct a challenging labeling task of choosing labels from a large candidate set. For example, in a recent semantic type detection method Sato~\cite{DBLP:journals/pvldb/ZhangSLHDT20}, there are in total 78 possible types which can make the labeling task quite difficult.


In \system, we take a different approach in the form of \emph{column matching} \revicde{to solve the semantic type detection task}. \system finds pairs of columns that have the same semantic type from a large collection of tables (e.g., the WebTable corpora \cite{DBLP:journals/pvldb/CafarellaHWWZ08}). These identified pairs can form clusters of same-type columns for which domain experts can easily assign the type by inspecting a few elements. This problem definition is closely related to the task of detecting pairs of related table columns~\cite{DBLP:journals/pvldb/LimayeSC10} and domain discovery of table values~\cite{DBLP:journals/pvldb/OtaMFS20}.
\revicde{Unlike existing methods~\cite{DBLP:conf/wsdm/ZhangWSDFP20,DBLP:conf/kdd/HulsebosHBZSKDH19}, \system performs column matching thus it does not depend on pre-defined semantic types.
\system is also self-supervised requiring significantly less labeling effort.}
	

We can adapt \system to this task as follows. Here, each data item corresponds to a table column
which can be serialized by concatenating the column values, for example: 

\begin{small}
\begin{center}
[VAL] New York [VAL] California [VAL] Florida
\end{center}
\end{small}

\noindent
similarly to EM where ``[VAL]'' indicates the start of a new column value. 
Note that to illustrate the effectiveness of \system,
we choose the bare-bone serialization method that does not take meta-information such as column names, adjacent columns,
or table descriptions as input. The meta-information can be integrated into this process by modifying the serialization scheme, for example, 
by adding the column name to the left of the sequence.

We also need to adjust the DA operators before applying the pre-training algorithm of \system here.
The attribute-level operators no longer apply.
In addition to the remaining token and span-level operators, we introduce a cell-level operator that shuffles the order
of the column values. 
We can apply the rest of components in \system without modification. 

Finally, \system outputs clusters of table columns that are inter-connected as the pairwise matching model $M_\pmm$ predicts. By adjusting the column clustering algorithm,
\system can generate clusters of different granularity
levels and identify semantic types that are not previously
defined in the multi-class classification setting.

\section{Experiments}\label{sec:experiment}

We evaluate the performance of \system on real-world datasets of entity matching (EM) and data cleaning.
Specifically, for EM, we evaluate the performance of both the blocking and matching stages. 
For data cleaning, we combine the error detection and correction stages and evaluate the quality of the final corrections.
We also conduct a case study of column semantic type discovery on the VizNet dataset. 
\revicde{Due to space limitation, we provide more results on running time, parameter sensitivity, data profiling, and error analysis in the appendix of the technical report ~\cite{DBLP:journals/corr/abs-2207-04122}.}

\subsection{Experiment settings}


For EM, we use 5 datasets 
provided by \dm~\cite{DBLP:conf/sigmod/MudgalLRDPKDAR18} which are widely used in previous studies.
These datasets are for training and evaluating EM models for various domains including products, publications, and businesses.
Each dataset consists of two entity tables A and B to be matched. Blocking methods take these 2 tables as input and
generate candidates of matched pairs.
For matching, each dataset provides sets of labeled entity pairs where each pair has a binary label
indicating whether it is a match / non-match.
The goal is to decide whether each pair represents the same real-world object, e.g. publication or product.
The original datasets are split into training, validation, and test sets 
at a 3:1:1 ratio.
Since \system targets application scenarios with insufficient label examples,
we consider the settings of \emph{semi-supervised} and \emph{unsupervised} EM
where \system only uses 500 or 0 labels from the training and validation set.
Table \ref{tab:em_dataset} shows the statistics of the datasets.

\setlength{\tabcolsep}{2.5pt}
\begin{table}[!t]
\caption{Statistics of EM datasets. }\label{tab:em_dataset}
\small
\begin{tabular}{cccccc} \toprule
Datasets            & TableA & TableB & Train+Valid & Test & \%pos   \\ \midrule
Abt-Buy (AB)        & 1,081   & 1,092   & 7,659        & 1,916 & 10.7\% \\
Amazon-Google (AG)  & 1,363   & 3,226   & 9,167        & 2,293 & 10.2\% \\
DBLP-ACM (DA)       & 2,616   & 2,294   & 9,890        & 2,473 & 18.0\% \\
DBLP-Scholar (DS)   & 2,616   & 64,263  & 22,965       & 5,742 & 18.6\% \\
Walmart-Amazon (WA) & 2,554   & 22,074  & 8,193        & 2,049 & 9.4\% \\ \bottomrule
\end{tabular}
\vspace{-3mm}
\end{table}

For data cleaning, we use 4 benchmark datasets provided by previous studies~\cite{DBLP:journals/pvldb/MahdaviA20,DBLP:conf/sigmod/MahdaviAFMOS019}.
Each dataset consists of a dirty spreadsheet and the goal is to generate a correction for each erroneous cells
(Error Correction, EC). A related task is to identify cells that contain errors, called Error Detection (ED).
Following the settings of \baran,
\system uses the same external EC modules for generating candidate corrections for each clean/dirty cell.
\system then utilizes the methods introduced in Section~\ref{sec:cleaning} to obtain the pairwise model for matching cells with their candidate corrections.
The details of datasets are shown in Table~\ref{tab:cleaning_dataset}.

\setlength{\tabcolsep}{2pt}
\begin{table}[!ht]
\caption{Statistics of data cleaning datasets. The error types are missing value (MV), typo (T), formatting issue (FI),
and violated attribute dependency (VAD)~\cite{DBLP:journals/pvldb/MahdaviA20}. 
\rev{*Note that the tax dataset is sampled from the original dataset of 200k rows.}}\label{tab:cleaning_dataset}
\small
\begin{tabular}{cccccc} \toprule
Datasets & size & \%error & Error Types    &  Coverage & \#candidates \\ \midrule
beers    & 2,410 $\times$ 11        & 16\%    & MV, FI, VAD    & 94.9\%       & 63.4         \\
hospital & 1,000 $\times$ 20        & 3\%     & T, VAD         & 89.5\%       & 68.3         \\
rayyan   & 1,000 $\times$ 11        & 9\%     & MV, T, FI, VAD & 51.4\%       & 215.6         \\
tax*      & 5,000 $\times$ 15        & 4\%     & T, FI, VAD     & 92.7\%       & 1,442.3        \\ \bottomrule
\end{tabular}
\vspace{-3mm}
\end{table}

\subsubsection{Baseline methods}



We compare \system with 
SOTA EM methods \dm and \ditto and also 3 specialized methods for semi-supervised
or unsupervised EM:

\smallskip
\noindent
\textbf{\dm}~\cite{DBLP:conf/sigmod/MudgalLRDPKDAR18} is a hybrid deep learning method based on
Recurrent Neural Network (RNN) and the attention mechanism. We report the scores from the original paper.

\smallskip
\noindent
\textbf{\ditto}~\cite{ditto2021} is the SOTA matching solution based on pre-trained LMs.
\ditto has a series of optimizations including data augmentation (DA).
We choose RoBERTa~\cite{DBLP:journals/corr/abs-1907-11692} as the base LM for \ditto which was shown to achieve the best performance.





\smallskip
\noindent
\textbf{\rot}~\cite{DBLP:conf/sigmod/Miao0021} is a semi-supervised approach that leverages
meta-learning to combine augmentations from multiple DA operators. \rot was shown to achieve the
SOTA results on EM in the low-resourced setting. 

\smallskip
\noindent
\textbf{\zeroer}~\cite{DBLP:conf/sigmod/WuCSCT20} is an unsupervised EM solution based on
Gaussian Mixture Models (GMM) for learning match/unmatch distributions.

\smallskip
\noindent
\textbf{\autofj}~\cite{DBLP:conf/sigmod/LiCCHC21} is an unsupervised approach that
automatically builds fuzzy join programs. This method relies on the assumption that
either Table A or B is a reference table with no or few duplicates.

For the blocking stage of EM, we compare \system with \textbf{\dbl}~\cite{DBLP:journals/pvldb/ThirumuruganathanLTOGPFD21}, 
a recently proposed deep learning framework achieving state-of-the-art (SOTA) results on blocking. \dbl follows \dm and leverages
a variety of deep learning techniques including self-supervised learning.

For data cleaning, we compare \system with the EC method
\textbf{\baran}~\cite{DBLP:journals/pvldb/MahdaviA20}. 
\baran leverages active learning to learn an ensemble model of multiple external EC methods.
The original setting of Baran assumes a perfect ED output or uses Raha~\cite{DBLP:conf/sigmod/MahdaviAFMOS019}
as the separate ED stage. 

\rev{
For column matching, we consider two
SOTA baseline methods 
\textbf{\sherlock}~\cite{DBLP:conf/kdd/HulsebosHBZSKDH19} and 
\textbf{\sato}~\cite{DBLP:journals/pvldb/ZhangSLHDT20}.
Both methods use ML/DL such as word2vec and LDA to represent
columns as dense vectors.}

For all the above baseline methods, we use the source code from the original code repositories to produce the results.





\subsubsection{Environment and hyper-parameters}

We implemented \system using PyTorch~\cite{DBLP:conf/nips/PaszkeGMLBCKLGA19} and Huggingface~\cite{DBLP:conf/emnlp/WolfDSCDMCRLFDS20}.
Unless otherwise specified, 
we use the RoBERTa-base model~\cite{DBLP:journals/corr/abs-1907-11692} as the pre-trained LM and AdamW as the optimizer for all the experiments.
We fix the projector dimension to 768 and the pre-training related 
hyper-parameters $(\lambda, \tau)$ to (3.9e-3, 0.07).
We pre-trained each model using 3 epochs and fine-tuned it for 50 epochs. We fix the size of the pre-training corpus to 10,000
by up or down-sampling the set of all data items.
We set the batch size to 64 and the learning rate to 5e-5.
The matching and data cleaning tasks use the F1 score as the main evaluation metric. 
\rev{
We list the hyper-parameters related to \system's optimizations in Table \ref{tab:hyperparameters}.
}
For each run of the experiment, we select the epoch with the highest F1 on the validation set and report results on the test set.
All experiments are run on a server machine with a configuration similar to 
a p4d.24xlarge AWS EC2 machine with 8 A100 GPUs.

\begin{table}[!ht]
\centering
\scriptsize
\caption{Hyper-parameters of \system and their choices.
We underlined the best combination found via grid search.}
\label{tab:hyperparameters}
\begin{tabular}{ccc}
\toprule
Hyper-Parameter & Meaning                                         & Range  \\ \midrule
cutoff\_ratio   & \%tokens to apply the cutoff DA                 & {[}0.01, 0.03, \underline{0.05}, 0.08{]}                                                                            \\
num\_clusters   & Number of clusters for neg. sampling        & {[}30, 60, \underline{90}, 120{]}                                                                                \\
alpha\_bt       & Weight of the redundancy regularizer            & 
\{0.1, 0.01, \underline{1e-3}, 1e-4\}
\\
multiplier      & Size of the training set w. pseudo labels & {[}2, 4, 6, \underline{8}, 10{]} \\ \bottomrule                                                
\end{tabular}
\vspace{-2mm}
\end{table}

\subsection{Main results for Entity Matching}

\setlength{\tabcolsep}{3.5pt}
\begin{table}[t]
    \centering
	\footnotesize
	\caption{\small F1 scores for semi-supervised matching (EM). 
\system uses 500 uniformly sampled pairs from train+valid.}\label{tab:semis}
\scriptsize
	\begin{tabular}{ccccccc}
		\toprule
		 & AB & AG & DA & DS & WA & average  \\
		\midrule
\dm(full) & 62.8 & 69.3 & 98.4 & 94.7 & 67.6 & 78.6 \\
\ditto(500)      & 70.1 & 44.7 & 95.9 & 89.4 & 49.4 & 69.9 \\
\ditto(750)      & 79.4 & 56.3 & 96.3 & 90.4 & 65.7 & 77.6 \\
\rot(500)      & 69.7 & 54.0 & 95.9 & 91.9 & 50.1 & 72.3 \\
\rot(750)      & 80.1 & 57.8 & 95.7 & 90.3 & 68.5 & 78.5 \\ 

SimCLR & 65.7 & 43.7 & 96.0 & 89.0 & 41.1 & 67.1 \\
\system(-cut,-RR,-cls)      & 75.8 & 56.8 & 95.0 & 90.6 & 65.6 & 76.7 \\
\system(-cut,-RR)            & 81.0 & 58.6 & 95.1 & 91.1 & 62.8 & 77.7 \\
\system(-cut)                 & 81.7 & 58.8 & 95.3 & 91.3 & 62.8 & 78.0 \\
\system(-PL)                  & 73.9 & 41.8 & 96.0 & 90.2 & 40.6 & 68.5 \\
\system(-RR)                  & 79.6 & 59.1 & 95.3 & 91.1 & 64.3 & 77.9 \\
\system(-cls)                 & 74.8 & 55.0 & 95.0 & 90.6 & 65.5 & 76.2 \\
\midrule
\system                        & 81.1 & 59.3 & 95.2 & 89.9 & 66.1 & 78.3 \\

$\Delta_1$ (vs. \rot, 500) & \green{(+11.4)} & \green{(+5.3)} & \red{(-0.7)} & \red{(-2.0)} & \green{(+16.0)} & \green{(+6.0)} \\
$\Delta_2$ (vs. SimCLR) & \green{(+15.4)} & \green{(+15.6)} & \red{(-0.8)} & \green{(+0.9)} & \green{(+25.0)} & \green{(+11.2)} \\
\bottomrule
	\end{tabular}
\end{table}


For semi-supervised learning, we set the label budget to 500 for \system.
We use the same 500 labels for validation for further label saving.
For the baselines \ditto and \rot, we allow for 250 more training instances (in total 750) 
and compare the results with \system with 500 labels to show the label efficiency of \system.
We fix the DA operator to be token\_del with span cutoff.
For pseudo-labeling, we tuned the positive/negative thresholds 
$(\theta^+, \theta^-)$ and found that adding 7x extra 
labels (i.e., multiplier=8)
works the best with the clustering optimization.
We fix the number of fine-tuning steps unchanged when adding the extra labels.


We also conduct an ablation analysis to show the effectiveness of each optimization.
\rev{We remove each of the following optimizations at a time
to test their overall effectiveness:
\begin{itemize}
\item \textsf{PL}: the pseudo labeling technique (Section~\ref{sec:pseudo});
\item \textsf{Cls}: the clustering-based negative sampling (Section~\ref{sec:clustering});
\item \textsf{Cut}: the cutoff operator for DA (Section~\ref{sec:cutoff});
\item \textsf{RR}: the redundancy regularization technique (Section~\ref{sec:barlow_twins}).
\end{itemize}
We denote each variant by ``\system (-X, -Y, \dots)''
indicating that optimizations X, Y, etc. are turned off.
Note that SimCLR~\cite{DBLP:conf/icml/ChenK0H20} 
is equivalent to \system with all 4 optimizations turned off.}

Table \ref{tab:semis} summarizes the results under the setting of semi-supervised learning.
\rev{When using 500 labels only,
\system achieves the overall best results and has a performance gain over \rot (500) by up to 16\%.
While \rot is already a label-efficient solution, 
\system can achieve comparable F1 scores (78.3 vs. 78.5)
by using 1/3 fewer labels (vs. \rot (750)). 
This F1 score is also close to \dm (full) 
using 23x more labels.
}

Pseudo-labeling (PL) is the most effective optimization
among the 4 tested options as removing it causes near 10\%
performance degradation. The 3 other optimizations,
Cls, Cut and, RR, are especially effective on the 
more challenging WA dataset. All optimizations combined
achieved up to 25\% F1 improvement on the WA dataset
or 11.2\% on average.

\smallskip
\noindent
\textbf{Unsupervised EM. }
Table~\ref{tab:uns} shows results of unsupervised EM.
Note that for pseudo labeling, \system requires prior knowledge of the positive label ratio
which is available as a dataset statistics. \system requires no more supervision beyond it.
\rev{
The results for \zeroer and \autofj are 
numbers either taken from the original papers or obtained by running the open-sourced code, 
whichever is higher.}

\rev{Compared to the two SOTA methods, \system outperforms \zeroer by up to 22.8\% (or 7.7\% on average)
and \autofj 
by up to 34.8\% (or 8.9\% on average) 
F1 score respectively.}
The average performance of the unsupervised \system is only 4\% lower than that with 500 labels.
\system's optimizations are again beneficial as they improve the base performance by 0.9\% on average. 
The main source of the performance gain is again from the high-quality pseudo labels
which are only slightly worse ($<$6\%) than the semi-supervised setting. 
These results further illustrate the advantage of \system 
in the low-resource scenarios.



\setlength{\tabcolsep}{5pt}
\begin{table}[t]
    
	\small
	\centering
	\caption{\small F1 scores for unsupervised matching (EM). 
	}\label{tab:uns}
	\footnotesize
	\begin{tabular}{ccccccc}
		\toprule
		& AB & AG & DA & DS & WA & average  \\ \midrule
		\zeroer &  52.0 & 48.0 & 96.0  & 86.0  & 51.0 & 66.6 \\
		\autofj & 61.3 & 24.3 & 91.6 & 88.5 & 61.3 & 65.4 \\
		\midrule
		\revicde{Sudowoodo (-cut,-RR,-cls)} &  75.2 & 53.5  & 90.2 & 85.0 & 63.3 & 73.4 \\
		\system & 74.8 & 59.1 & 91.6 & 87.1 & 61.2 & 
		\textbf{74.3} \\
		\bottomrule
	\end{tabular}
	\vspace{-4mm}
\end{table}

\begin{figure*}[!t]
    \centering
    \includegraphics[width=0.9\textwidth]{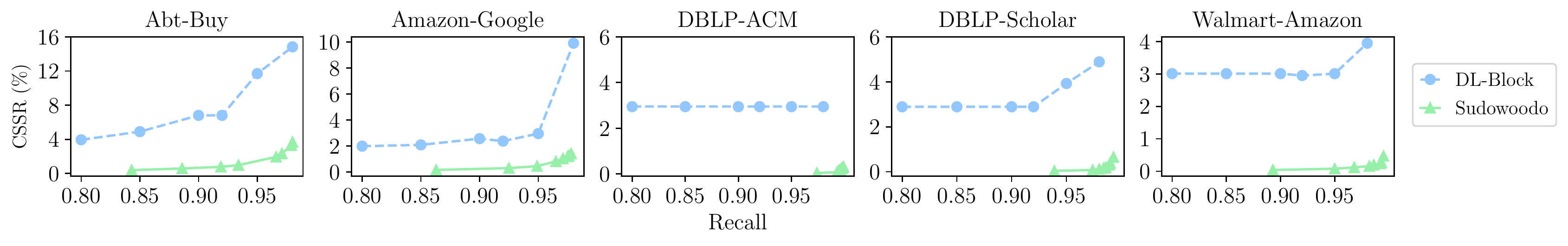}
    \caption{\small Recall-CSSR (candidate set size ratio) curves of \system vs. \dbl. 
    Compared to \dbl, \system produces candidates of better quality as it has both higher recall and lower CSSR (bottom right).
    The numbers for \dbl are from Figure 6 of \cite{DBLP:journals/pvldb/ThirumuruganathanLTOGPFD21}.}
    \label{fig:blocking}
	\vspace{-2mm}
\end{figure*}
\smallskip
\noindent
\textbf{EM Blocking. } 
For blocking, \system generates candidate pairs by
applying kNN search over 
the learned vector representations of
the right entity table (Table B)
for $k=1$ to 20. We find \system achieves
the best results using the
DistilBERT~\cite{DBLP:journals/corr/abs-1910-01108} LM
and a projector size of 4,096. We pre-train \system for 5 epochs
and use the last model checkpoint to generate candidates for evaluation.

We compare the quality of the candidate sets with the SOTA learning-based blocking method \dbl.
Following \cite{DBLP:journals/pvldb/ThirumuruganathanLTOGPFD21},
we report the recall scores and the candidate set size ratio (CSSR) which equals the candidate set size
divided by the total number of candidates (i.e., $|$TableA$|$ $\times$ $|$TableB$|$).
For each dataset, we compute the recall score as the fraction
of positive pairs from the training, validation, and test sets 
that are retained by the candidate set.

Figure \ref{fig:blocking} shows the Recall-CSSR curves and
Table \ref{tab:block} shows the recalls and candidate set sizes of \dbl and \system. Here \dbl's numbers 
are taken from \cite{DBLP:journals/pvldb/ThirumuruganathanLTOGPFD21}.
In Table \ref{tab:block}, 
we report \system's (recall, \#cand) for the first $k$ where 
\system's recall is higher than \dbl.
The results show that \system can consistently achieve the same recall level of \dbl while
having a significantly smaller candidate set (up to 84.8\% smaller, AB).
By having a smaller candidate set, \system reduces the difficulty of the matching stage 
which can lead to further 
matching performance improvement or 
saving of human efforts.
These results confirm \system's versatility in
benefiting multiple sub-tasks of the EM application.

\revicde{We provide ablation study, running time, data profiling and error analysis in Appedix A, B and E respectively in our technical report version \cite{DBLP:journals/corr/abs-2207-04122} }

\setlength{\tabcolsep}{2.3pt}
\begin{table}[ht]
	\small
	\caption{\system for blocking. We report the recall score (R) and the number of candidates pairs (\#cand).}\label{tab:block}
\resizebox{0.48\textwidth}{!}{
\begin{tabular}{c|cccccccccc}\toprule
          & \multicolumn{2}{c}{AB} & \multicolumn{2}{c}{AG} & \multicolumn{2}{c}{DA} & \multicolumn{2}{c}{DS} & \multicolumn{2}{c}{WA} \\
          & R         & \#cand     & R         & \#cand     & R         & \#cand     & R         & \#cand     & R         & \#cand     \\ \midrule
DL-Block  & 87.2      & 21,600      & 97.1      & 68,200      & 99.6      & 13,100      & 98.1      & 392,400     & 92.2      & 51,100     \\ 
\system & \textbf{88.6}      & \textbf{3,276}       & \textbf{97.3}      & \textbf{48,390}      & 99.6      & \textbf{11,470}      & \textbf{98.4}      & \textbf{257,052}     & \textbf{95.0}      & \textbf{44,148}      \\ \bottomrule
\end{tabular}}
\vspace{-6mm}
\end{table}

\subsection{Results for data cleaning}\label{subsec-dcexp}

We evaluate \system on data cleaning
comparing it with the SOTA EC solution \baran~\cite{DBLP:journals/pvldb/MahdaviA20}.
We run \system with all optimizations turned on except pseudo labeling.
We use the span\_shuffle DA operator with span cutoff for pre-training.
We did not use blocking since the candidate correction sets are small (see Table \ref{tab:cleaning_dataset}).
After pre-training, we fine-tuned the representation model on \rev{20 uniformly sampled rows (same amount of
supervision for \baran using active learning).
We evaluate the F1 score of \system on the remaining rows
following the same setting of \baran.}


Table \ref{tab:dc} summarizes the results.
\baran has two settings: Raha+\baran uses Raha~\cite{DBLP:conf/sigmod/MahdaviAFMOS019} as the ED model 
run before \baran. The second setting (Perfect ED + \baran) assumes an ED model that perfectly identifies all the dirty cells.
We also report a method RoBERTa-base which fine-tunes the LM without \system's pre-training step.

\system achieves the best overall results even outperforming
\baran with perfect ED by 2.1\% on average.
\system outperforms Raha+\baran by a large margin and
in 3 out of 4 datasets by up to 44\% F1 score (hospital).
The results indicate that \system's strategy of formulating EC as a matching problem is beneficial.
The contrastive pre-training is also important because without this step,
RoBERTa LM has a performance degradation 
of up to 15.6\% (hospital) F1 score.
These results show the potential of integrating \system into a data cleaning pipeline in real-world scenarios.

\revicde{We provide detailed ablation results
 of \system on EC in Appendix \ref{app:dc}  of our technical report version\cite{DBLP:journals/corr/abs-2207-04122}.}

\begin{table}[ht]
	\small
	\centering
	\caption{Error correction (EC) F1 scores for data cleaning. \system does not rely on a separate Error detection (ED) stage. }\label{tab:dc}
	\begin{tabular}{cccccc}
		\toprule
		& beers & hospital & rayyan & tax  & average  \\
		\midrule
		Raha + \baran & 95.74 & 45.32 & 46.10 & 70.07 & 64.31 \\
		Perfect ED + \baran & 90.43 & {90.48} & 63.51 & 80.73 & 81.29 \\
		RoBERTa-base & {90.75} & 73.33 & 61.83 & 87.62 & 78.38 \\
		\system & 90.71 & 89.01 & {62.91} & {91.22} & \textbf{83.46} \\
		\bottomrule
	\end{tabular}
	\vspace{-4mm}
\end{table}

\subsection{Results for \revicde{semantic type detection}}\label{sec:coltype}

We conduct a case study where \system is applied to the column matching task on the VizNet dataset.
Following the setting in \cite{DBLP:journals/pvldb/ZhangSLHDT20}, we extracted more than 80k tables with 119k columns 
from the dataset annotated with 78 semantic types.
We apply the column matching approach described in Section \ref{sec:column}.
We use kNN blocking with $k=20$ to extract pairs of nearest columns as candidate matches.
After blocking, we use the ground truth labels 
from the original dataset to label 2k uniformly sampled pairs 
for training where we labeled a pair to be ``match'' 
if and only if they have the same original semantic type.
We split this dataset into the training/validation/test sets according to the 2:1:1 ratio.

\begin{table*}[!ht]
	\caption{Column clusters discovered by \system. 
		The shown values are the first elements of 5 randomly selected columns having the same type.
		The headers (name, club, \dots) are the majority ground truth types
		from~\cite{DBLP:conf/kdd/HulsebosHBZSKDH19,DBLP:journals/pvldb/ZhangSLHDT20}.
		\system discovers fine-grained types 
		(e.g., central EU city)
		beyond the original set of 78 labels.
		We also provide our interpretations of such clusters/types (the term after the ``$\rightarrow$''). }
	\label{tab:column}
	\small
	\resizebox{0.98\textwidth}{!}{
		\begin{tabular}{c|c|c|c|c|c|c|c|c} \toprule
			name           & club & language           & \begin{tabular}{c} state $\rightarrow$ \\  US State \end{tabular} & 
			\begin{tabular}{c} result $\rightarrow$ \\  ball game result \end{tabular}   & 
			\begin{tabular}{c} name $\rightarrow$ \\  company name \end{tabular}                           & weight         & 
			\begin{tabular}{c} city $\rightarrow$ \\  central EU city \end{tabular}      & 
			\begin{tabular}{c} result $\rightarrow$ \\  baseball in-game event \end{tabular}                       \\ \midrule
			Geerhart, Kyle   & AMS  & Afrikaans          & TX    & Win      & Lone Pine Capital LLC          & 50 lbs or less & Marburg   & single, left field                       \\
			Hossfeld, Nick  & SDSM & Polski             & NV    & Win, 3-1 & T. Rowe Price Associates, Inc. & 38kg           & Stollberg & pop fly out, center field                  \\
			Dege, Henry & GAKW & Spanish            & LA    & 3-1 L    & Trigran Investments, Inc.      & 40 lbs         & Pratteln  & strikeout                    \\
			Carlisle, Brenden   & WSM  & English (built-in) & AZ    & W 9-0    & Icahn Associates Corp.         & up to 25 lbs   & Berlin    & pitcher to first base \\
			Tatlow, Jedidiah & DCM  & Turkish            & NJ    & Win      & Apple Inc.                     & 5 lbs          & Osnabruck & walk             \\ \bottomrule
	\end{tabular}}
	\vspace{-3mm}
\end{table*}

\rev{
We compare the matching quality of \system with those of
the SOTA methods, \sherlock and \sato. Table \ref{tab:coltype}
shows the detailed results. Here, we use \sherlock and \sato
to extract features of the input pairs of columns
for training pairwise matching models
of Logistic Regression (LR), SVM, Random Forest (RF), and Gradient
Boosting Tree (GBT). Among the 4 models, GBT achieves the best
F1 on the validation set. 
The results show that \system outperforms
both \sherlock and \sato with GBT classifiers 
by  4.5\% and 3.7\% respectively.
}

\rev{
Using the learned column embedding and matching models,
\system discovers over 5k column clusters (semantic types) from the table corpus.
By inspecting the results, we find that \system can discover more fine-grained types 
that are not present in the 78 ground truth types (Table \ref{tab:column}).
Note that \system achieves this result by using only 1k column pair labels instead of 
more than 80k multi-class single-column labels required by \sato or \sherlock.}

\revicde{We provide more detailed experiments in the technical report version (see Appendix \ref{app:ctd}) to compare different methods in our technical  report version\cite{DBLP:journals/corr/abs-2207-04122}.}

\setlength{\tabcolsep}{3.5pt}
\begin{table}[!t]
\small
\centering
\caption{Results for column matching.} \label{tab:coltype}
\begin{tabular}{ccccccc} \toprule
             & \multicolumn{3}{c}{Valid}  & \multicolumn{3}{c}{Test}   \\
             & Precision & Recall & F1    & Precision & Recall & F1    \\ \midrule
\sherlock & 81.49     & 82.23  & 81.86 & \textbf{85.50}     & 82.34  & 83.89 \\
\sato     & \textbf{83.69}     & 83.43  & 83.56 & 84.57     & 84.33  & 84.45 \\
\system    & {83.62}     & \textbf{89.16}  & \textbf{86.30} & 85.37     & \textbf{91.45}  & \textbf{88.31} \\ \bottomrule
\end{tabular}
\vspace{-4mm}
\end{table}

\section{Related Work}\label{sec:related}


\noindent
\textbf{Data cleaning. } Deep learning has recently achieved great success in data cleaning and integration~\cite{DBLP:journals/pvldb/DongR18}.
While traditional approaches for data cleaning employ rule-based~\cite{DBLP:conf/sigmod/BohannonFFR05,DBLP:conf/icdt/KolahiL09,DBLP:journals/pvldb/AbedjanAOPS15} or statistical~\cite{DBLP:journals/pvldb/RekatsinasCIR17} methods, machine learning is now playing an increasingly more important role.
\rev{\textsf{ActiveClean}~\cite{DBLP:journals/pvldb/KrishnanWWFG16} employs active learning techniques for data cleaning.}
For error detection, 
\textsf{HoloDetect}~\cite{DBLP:journals/pvldb/RekatsinasCIR17,DBLP:conf/sigmod/HeidariMIR19} enriches the training data with task-specific data augmentation techniques under the setting of few-shot learning.
\textsf{Raha}~\cite{DBLP:conf/sigmod/MahdaviAFMOS019} improves the training process by leveraging advanced selection techniques for representative values.
\textsf{HoloClean}~\cite{DBLP:journals/pvldb/RekatsinasCIR17}
is a qualitative approach that generates probabilistic correction programs.
\textsf{Baran}~\cite{DBLP:journals/pvldb/MahdaviA20} solves error correction
by utilizing the ensemble learning method to combine multiple error correction modules.
See~\cite{DBLP:conf/edbt/Thirumuruganathan20} for a detailed survey
that summarizes future directions for adopting ML for data cleaning.

\smallskip
\noindent
\textbf{Entity Matching. }
Entity Matching (EM) is an important data integration task
that has been extensively studied over the past decades~\cite{DBLP:journals/pvldb/GetoorM12,DBLP:journals/pvldb/KondaDCDABLPZNP16}.
There are two main steps in an EM pipeline: blocking and matching.
The blocking step aims at reducing the number of potential comparisons 
from the quadratic-size all possible pairs.
The goal is to generate a small candidate set while retaining 
as many real matches as possible.
Example techniques include rule-based blocking~\cite{DBLP:conf/sigmod/GokhaleDDNRSZ14,DBLP:conf/sigmod/DasCDNKDARP17}, schema-agnostic blocking~\cite{DBLP:journals/tkde/SimoniniPPB19}, meta-blocking~\cite{DBLP:journals/pvldb/SimoniniBJ16} and deep learning approaches~\cite{DBLP:conf/wsdm/ZhangWSDFP20,DBLP:journals/pvldb/ThirumuruganathanLTOGPFD21}.
The matching step performs the pairwise comparisons to identify matched entity entries, 
where deep learning-based techniques are currently achieving promising results.
\textsf{DeepER}~\cite{DBLP:journals/pvldb/EbraheemTJOT18} and \dm~\cite{DBLP:conf/sigmod/MudgalLRDPKDAR18} employ the Recurrent Neural Network (RNN) models to perform entity matching.
Kasai et al.~\cite{DBLP:conf/acl/KasaiQGLP19} develops the active learning method for EM with insufficient labeled examples.
\textsf{Seq2SeqMatcher}~\cite{DBLP:conf/cikm/NieHHSCZWK19} and \textsf{HierMatcher}~\cite{DBLP:conf/ijcai/FuHHS20} improve the performance of matching between heterogeneous data sources by applying additional alignment layers. 
\textsf{Rotom}~\cite{DBLP:conf/sigmod/Miao0021} utilizes meta-learning 
and data augmentation techniques to support data integration tasks 
including EM.
Some recent studies~\cite{DBLP:conf/edbt/BrunnerS20,DBLP:conf/vldb/PeetersBG20,ditto2021} apply pre-trained language models such as BERT for EM. Unlike these recent methods that optimize one component at a time, 
\system provides a solution achieving promising results in both blocking and matching.

\smallskip
\noindent
\textbf{Semantic type detection. }
Semantic type detection refers to the task of assigning semantics-rich types to table columns
to enhance functionalities of data exploration systems such as 
Microsoft Power BI~\cite{ferrari2016introducing} and 
Google Data Studio~\cite{snipes2018google}.
Previous methods include both ontology-based~\cite{DBLP:journals/pvldb/CafarellaHWWZ08,DBLP:conf/www/LehmbergRMB16,DBLP:conf/edbt/RitzeB17} and learning-based methods~\cite{DBLP:journals/pvldb/LimayeSC10,DBLP:conf/aaai/TakeokaONO19,DBLP:conf/kdd/HulsebosHBZSKDH19,DBLP:journals/pvldb/ZhangSLHDT20}.
In particular, Sato~\cite{DBLP:journals/pvldb/ZhangSLHDT20} formulates semantic type detection as a multi-class classification
problem with 78 pre-defined types and achieves the SOTA performance by leveraging conditional random field (CRF) and deep learning.
Unlike Sato, \system formulates the task as column matching and can detect fine-grained semantic types beyond the pre-defined
set of classes.

\smallskip
\noindent
\textbf{Contrastive learning}. Contrastive learning has been a popular and effective technique in self-supervised representation learning for many applications.
It is first applied in computer vision related tasks~\cite{DBLP:conf/cvpr/He0WXG20,DBLP:journals/corr/abs-2003-04297,DBLP:conf/nips/ChenKSNH20,DBLP:conf/icml/ChenK0H20,DBLP:conf/icml/ZbontarJMLD21} based on the intuition that good representations should be invariant under different distortions.
Contrastive learning has also been recently applied to NLP tasks 
by introducing a series of NLP-specific data augmentation operations.
\textsf{CERT}~\cite{DBLP:journals/corr/abs-2005-12766} extends the idea of MoCo~\cite{DBLP:conf/cvpr/He0WXG20} to utilize back-translation for data augmentation.
\textsf{CLUTR}~\cite{DBLP:conf/acl/GiorgiNWB20} borrows the idea from SimCLR~\cite{DBLP:conf/icml/ChenK0H20} to jointly train representations 
using a contrastive objective and masked language modeling.
\textsf{ConSERT}~\cite{DBLP:conf/acl/YanLWZWX20} adopts contrastive learning to fine-tune BERT model in an unsupervised way for multiple downstream tasks.
To the best of our knowledge, our work is the first one to employ contrastive learning
in data integration and preparation applications.
\vspace{-1mm}
\section{Conclusion}\label{sec:conclude}

We propose \system, a multi-purpose data integration and preparation (\di) framework based on contrastive learning.
It pre-trains high-quality representation models that can be fine-tuned to
downstream \di tasks using few or even no labels.
Unlike previous studies that optimize a single \di task at a time, 
we show that \system's representation models are applicable
to a range of \di tasks including Entity Matching (EM), data cleaning,
and data discovery.
Our experiment results show that \system achieves multiple SOTA results
under different supervision levels for both the blocking and matching 
step for EM. The promising results also extend
to the applications of data cleaning and column matching, which shows
\system's versatility.
For future work, we plan to study the integration of \system with other closely related learning paradigms such as distant supervision, active learning, and domain adaptation.

\bibliographystyle{IEEEtran}
\bibliography{main}
\balance

\newpage
\appendix

\subsection{Full ablation and parameter sensitivity analysis}\label{app:para}

\subsubsection{Detailed ablation analysis}

According to Table \ref{tab:semis},
among the 4 optimizations,
(1) Pseudo-Labeling (PL),
(2) clustering-based negative sampling (Cls),
(3) Cutoff data augmentation (Cut) and
(4) Redundancy Regularization (RR), 
PL is the most effective as removing it causes close to 10\%
of performance degradation on average.
We further verify the effectiveness of pseudo labeling
by checking the label quality. 
Table \ref{tab:pseudo} shows that after adding the pseudo labels,
the training set still has a high accuracy 
from 71\% to 99\% for positive labels
and 96\% to 99\% for negative labels.
The Cls and RR are effective among the rest techniques as they both achieved $>$2\% improvement on top of PL.

\setlength{\tabcolsep}{5pt}
\begin{table}[!ht]
\caption{True Positive Rate (TPR) and True Negative Rate (TNR) of 
the training set after adding pseudo labels.}\label{tab:pseudo}
\small
\centering
\begin{tabular}{ccccccc} \toprule
               & \multicolumn{2}{c}{SimCLR} & \multicolumn{2}{c}{\system} & \multicolumn{2}{c}{\system (no label)} \\
               & TPR          & TNR         & TPR           & TNR           & TPR               & TNR              \\ \midrule
AB        & 78.6         & 97.0        & 96.4          & 99.6          & 95.4              & 99.4             \\
AG  & 76.3         & 96.3        & 81.8          & 96.6          & 79.3              & 96.2             \\
DA       & 99.8         & 98.6        & 99.8          & 98.9          & 99.6              & 98.8             \\
DS   & 99.2         & 99.5        & 92.3          & 98.0          & 99.1              & 99.3             \\
WA & 69.4         & 97.0        & 71.7          & 97.0          & 66.0              & 96.5            \\ \bottomrule
\end{tabular}
\end{table}

{
The clustering-based negative sampling (cls) is also 
essential as it achieves up to 6.3\% gain of F1 score (AB).
The cutoff DA operator (cut) and redundancy regularization (RR) are especially useful in the WA dataset
achieving a 3.3\% and 1.74\% F1 improvement respectively.
All optimizations combined result in an F1 improvement
of up to 25\% (WA) or 11.2\% on average compared to the
base SimCLR self-supervised learning approach.
This significant gap shows the necessity of the
proposed optimizations techniques.}




{We note that \system does not consistently outperform 
the orthogonal meta-learning DA method \rot.
It is an interesting future work to explore how to 
integrate relevant fine-tuning methods like \rot
with SSL methods such as \system.}

\subsubsection{Hyper-parameter sensitivity}

\begin{figure*}[t]
    \centering
    \includegraphics[width=0.98\textwidth]{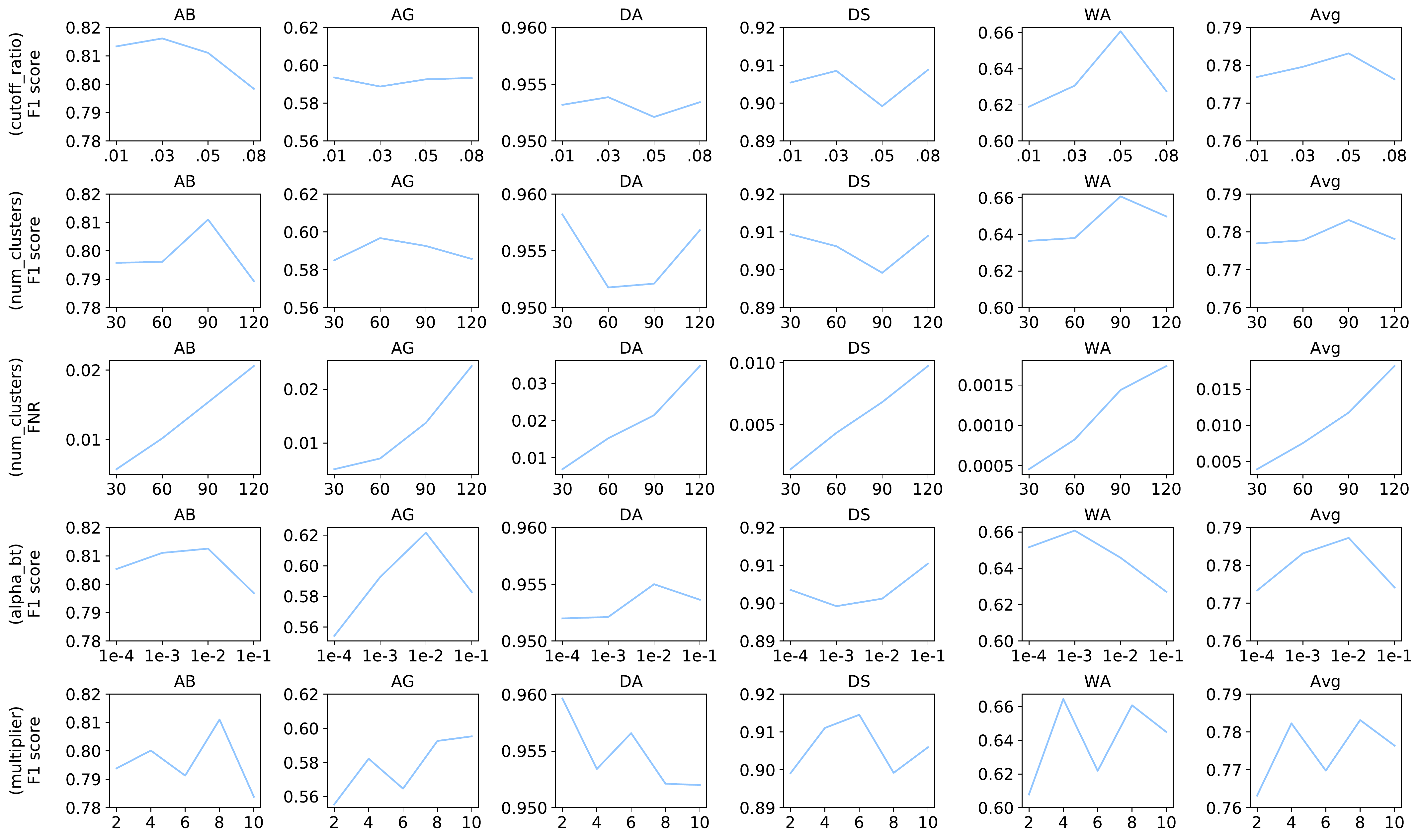}
    \caption{Parameter sensitivity analysis. The F1 scores of 
    \system is more stable when varying cutoff\_ratio and
    num\_clusters (1st and 2nd row). 
    \system is more sensitive to the alpha\_bt 
    and multiplier hyper-parameters (3rd and 4th row). }
    \label{fig:sensitivity}
\end{figure*}

The detailed results of parameter sensitivity analysis are shown in Figure~\ref{fig:sensitivity}.
Recall that we performed a grid search over 4 hyper-parameters
listed in Table \ref{tab:hyperparameters}:
\textsf{cutoff\_ratio}, 
\textsf{num\_clusters}, 
\textsf{alpha\_bt}, and 
\textsf{multiplier}
(see Table \ref{tab:hyperparameters}
for their definitions).
We have the following observations: 
Firstly, the performance of \system stays stable under most settings.
When varying the 4 hyper-parameters in the whole range
the average changes in the F1 
score are 0.41\%, 0.41\%, 0.57\%, and 0.52\%, respectively
(rows 1, 2, 4, and 5 of Figure \ref{fig:sensitivity}).
Secondly, we find that when we increase the number of clusters
for negative sampling, the false-negative ratio will increase linearly w.r.t the number of clusters, as expected (3rd row of Figure \ref{fig:sensitivity}). 
Nevertheless, the absolute number of false-negative tuples is 
within an acceptable rate of $<$2\% when the number of clusters
is within 90.
So one can avoid a high rate of false-negatives during
\system's pre-training by selecting a cluster size of about 1/100
of the number of unlabeled pairs.
Thirdly, the varying pattern of results w.r.t. ratio of pseudo labels is relatively complex. 
The users may need to select a proper value by grid search.


\subsection{Full results of running time}\label{app:time}

\begin{figure*}
\begin{minipage}{.36\textwidth} %
      \centering
      \includegraphics[width=1.0\textwidth]{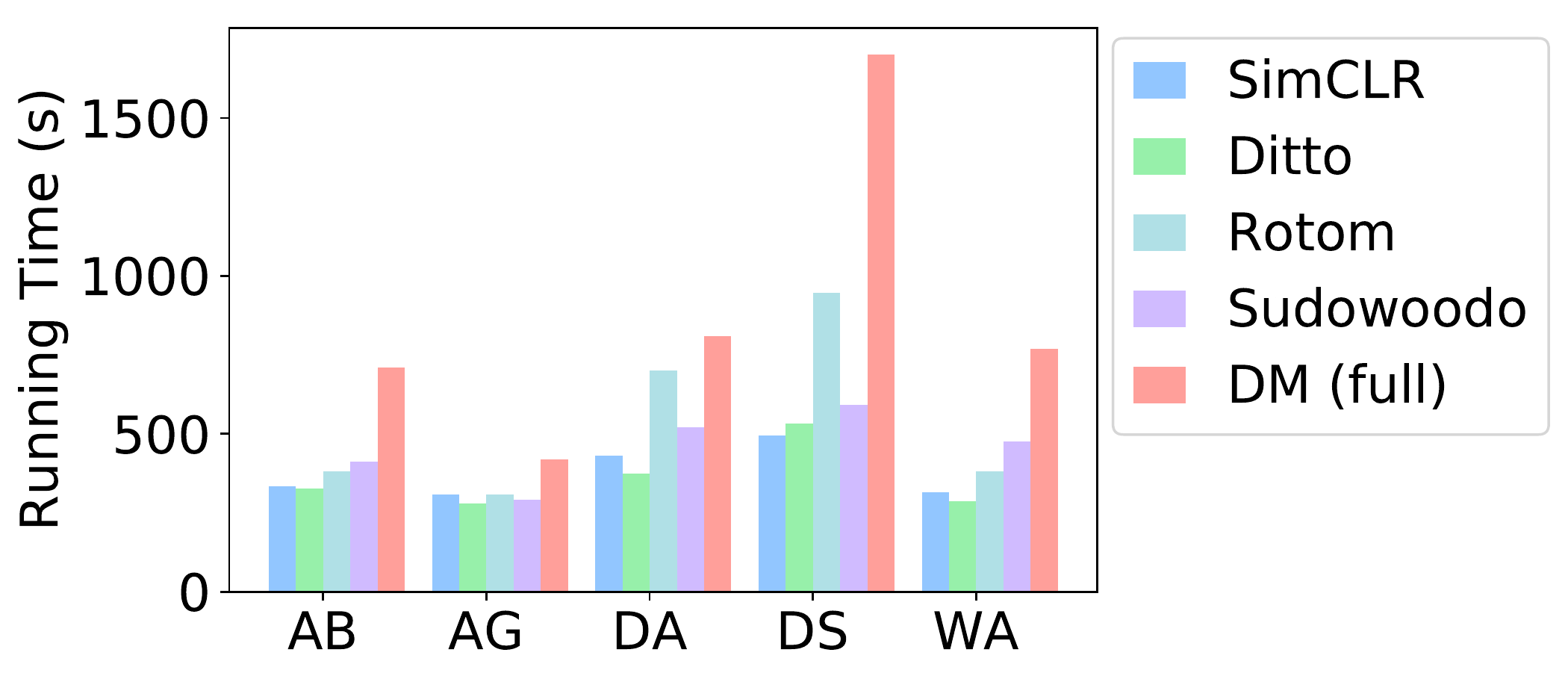}
      \caption{Running time for EM.}
      \label{fig:em_time}
\end{minipage} %
\begin{minipage}{.26\textwidth} %
      \centering
      \includegraphics[width=1.0\textwidth]{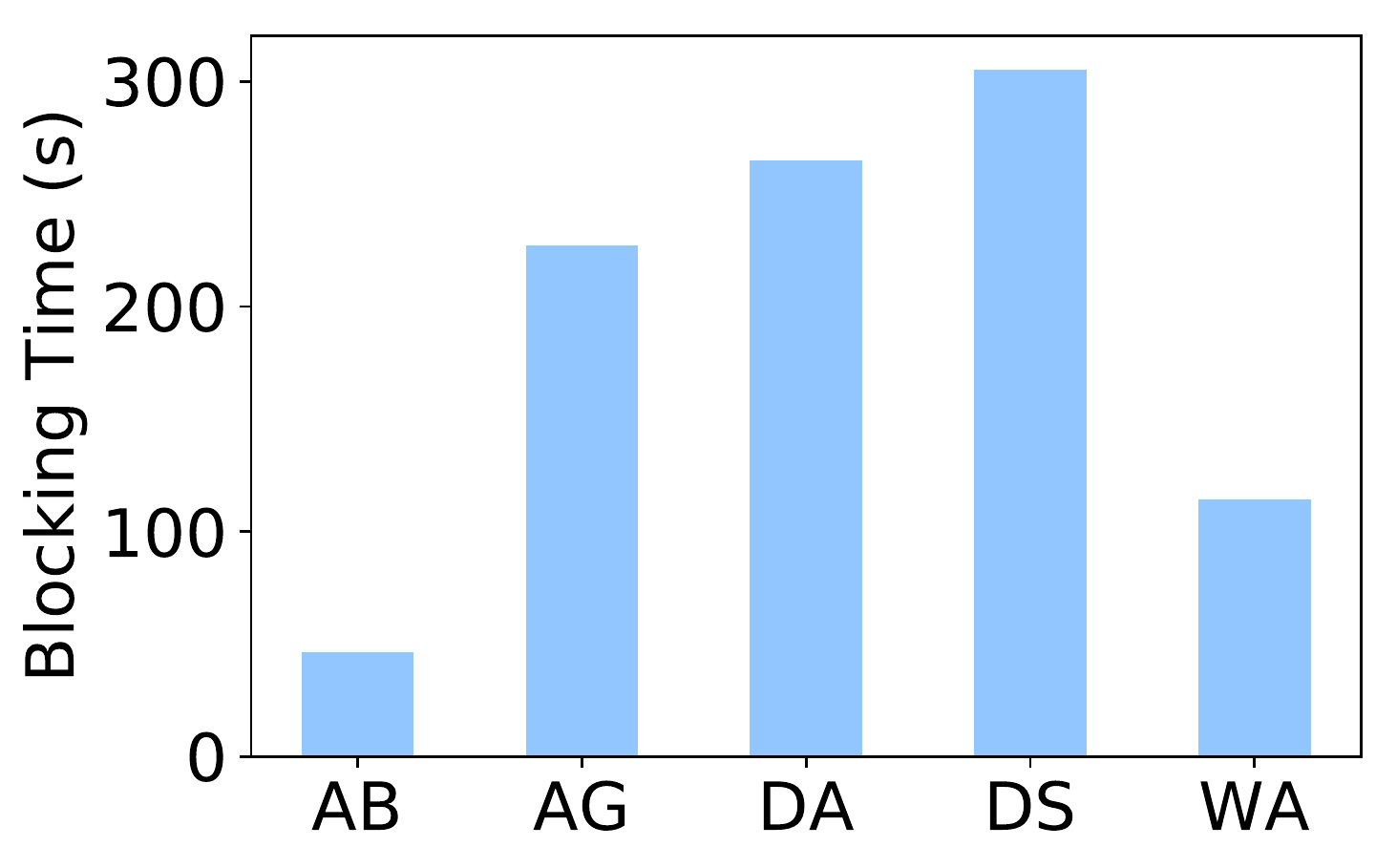}
      \caption{Blocking time for EM.}
      \label{fig:blocking_time}
\end{minipage}
\begin{minipage}{.36\textwidth} %
      \centering
      \includegraphics[width=1.0\textwidth]{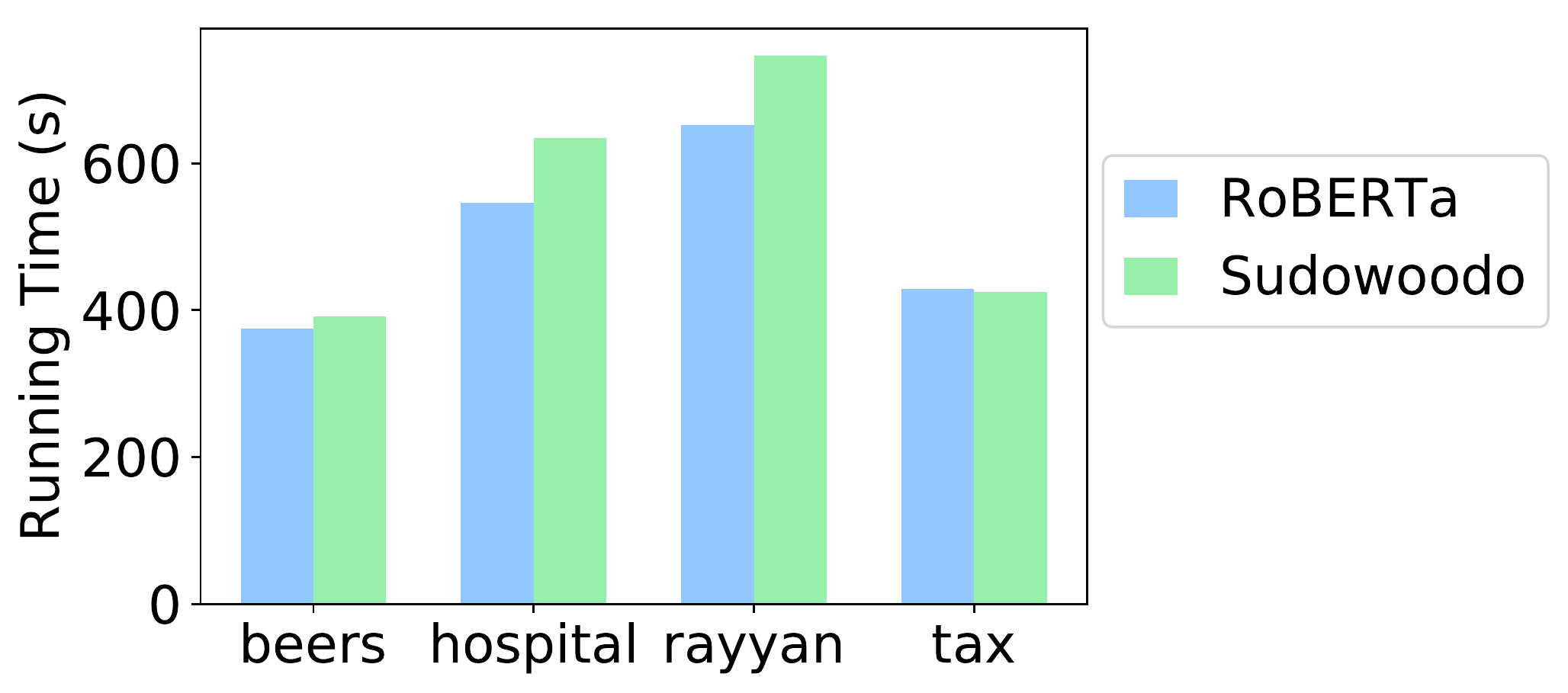}
      \caption{Running time for Data Cleaning.}
      \label{fig:dc_time}
\end{minipage}
\end{figure*}

We report the running time of \system on 3 applications: semi-supervised EM, blocking in EM, and data cleaning.
Figure~\ref{fig:em_time} shows the results of semi-supervised EM.
We can see that although \system involves extra steps of pseudo labeling, the performance is still comparable with the other pre-trained LM-based method \ditto.
Compared with \rot, \system is faster on 3 out of 5 datasets and the results are close on the remaining two datasets.
All methods are significantly faster than DeepMatcher training
on the full datasets.
Figure~\ref{fig:blocking_time} shows the results of blocking in EM. 
\system performs rather well in the blocking step: it takes less than 50 seconds on the AB dataset.
Even on the largest dataset DS, it only takes 305 seconds to finish.
Figure~\ref{fig:dc_time} illustrates the results of data cleaning.
We compare \system with the original RoBERTa language model.
It shows that the extra self-supervised learning steps
only add a small margin to the fine-tuning process.
This means that we can obtain a significantly better
pre-trained model at a relatively low cost.

\subsection{Full results of column type detection}\label{app:ctd}

Table \ref{tab:coltype_detailed} shows the performance
of \system versus Sherlock and Sato classifiers for the column pair
matching task. Both Sherlock and Sato were originally designed
to perform single-column classification. To apply them on
the pairwise column matching task, 
we use Sherlock and Sato as feature
extractors to represent each input column as a dense vector.
For a pair of columns $(c, c')$ with representations
$(\mathsf{vec}(c), \mathsf{vec}(c'))$, we train pairwise
classifiers $M$ using $$\mathsf{concat}\left(\mathsf{vec}(c), \mathsf{vec}(c'), |\mathsf{vec}(c) - \mathsf{vec}(c')|\right)$$
as input. Each input vector is of dimension
3,561 and 4,761 for Sherlock and Sato respectively.

We consider 5 widely used classification models: 
Logistic Regression (LR), Support Vector Machine (SVM),
Gradient Boosting Tree (GBT), Random Forest (RF), and
an additional baseline SIM that only uses the cosine similarity
$\mathsf{cos}(\mathsf{vec}(c), \mathsf{vec}(c'))$ as input.
By pairing Sherlock/Sato with the 5 classification models,
we obtain 2$\times$5$=$10 baseline variants.
Among all the variants, 
the GBT classifiers achieve the best F1 scores 
on the validation set. As we mentioned in Section \ref{sec:experiment}, \system outperforms all variants of 
Sherlock and Sato. The performance gain is mainly due to the
higher-quality column representations obtained via 
\system's pre-training steps.

\begin{figure*}[t]
	\centering
	\includegraphics[width=0.98\textwidth]{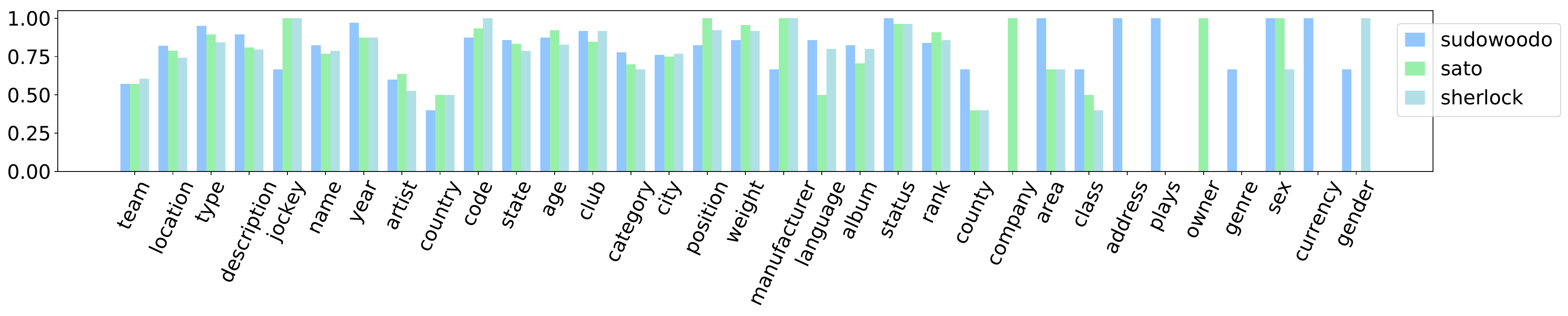}
	\caption{Column type detection results breakdown on fine-grained columns types.}
	\label{fig:coltype}
\end{figure*}

We further show the breakdown of Sato-GBT, Sherlock-GBT, and \system
on different ground truth types in Figure \ref{fig:coltype}.
\system outperforms Sherlock and Sato on popular types such as 
``location'', ``type'', and ``description''.
\system also performs well on less popular types such as
``area'', ``address'', ``plays'', and ``currency'' 
for which Sherlock and Sato fail to predict correctly.

\begin{table}[!ht]
\scriptsize
\centering
\caption{Full results of column type detection.
Each baseline classifier (LR, SVM, GBT, RF, and SIM) 
uses Sato/Sherlock column vectors as input features. }
\label{tab:coltype_detailed}
\begin{tabular}{ccccccc} \toprule
             & \multicolumn{3}{c}{valid}  & \multicolumn{3}{c}{test}   \\
             & Precision & Recall & F1    & Precision & Recall & F1    \\ \midrule
Sato-LR      & 84.19     & 73.80  & 78.65 & 86.85     & 80.91  & 83.78 \\
Sato-SVM     & 82.19     & 72.29  & 76.92 & 87.09     & 82.62  & \textbf{84.80} \\
Sato-GBT     & 83.69     & 83.43  & \textbf{83.56} & 84.57     & 84.33  & 84.45 \\
Sato-RF      & 84.47     & 78.61  & 81.44 & 81.47     & 78.92  & 80.17 \\
Sato-SIM     & 72.10     & 69.28  & 70.66 & 77.85     & 72.08  & 74.85 \\ \midrule
Sherlock-LR  & 82.51     & 75.30  & 78.74 & 83.68     & 80.34  & 81.98 \\
Sherlock-SVM & 81.58     & 74.70  & 77.99 & 83.33     & 76.92  & 80.00 \\
Sherlock-GBT & 81.49     & 82.23  & \textbf{81.86} & 85.50     & 82.34  & \textbf{83.89} \\
Sherlock-RF  & 82.03     & 75.60  & 78.68 & 84.71     & 82.05  & 83.36 \\
Sherlock-SIM & 70.96     & 71.39  & 71.17 & 74.13     & 72.65  & 73.38 \\ \midrule
Sudowoodo    & 83.62     & 89.16  & \textbf{86.30} & 85.37     & 91.45  & \textbf{88.31} \\ \bottomrule
\end{tabular}
\end{table}

\setlength{\tabcolsep}{3pt}
\begin{table}[!ht]
	\small
	\caption{Dataset and model statistics for column type detection.} \label{tab:column_stat}
	\resizebox{0.48\textwidth}{!}{
		\begin{tabular}{c|ccc|ccc}\toprule
			Dataset & \multicolumn{3}{c|}{Blocking} & \multicolumn{3}{c}{Matching}              \\ 
			\#columns                  & \#candidates                & \%pos                       &  Time & $|$Train$|$                       &  Time & \#clusters \\ \midrule
			\multicolumn{1}{c|}{119,360} & \multicolumn{1}{c}{1,526,555} & \multicolumn{1}{c}{67.95\%} & 12m58s    & 1,000 &  39m9s & 5,868 \\ \bottomrule
	\end{tabular}}
\end{table}

Table \ref{tab:column_stat} shows statistics of the 
blocking and matching steps on the VizNet column type detection dataset.
Using the learned embedding and matching models,
\system discovers over 5k column clusters
(semantic types) by applying a connected component algorithm 
to the pairwise matching results. 
The column clusters are high-quality as they achieve
an average purity of 89.9\%. 
By inspecting the results, we find that \system can discover more fine-grained
types compared to the 78 ground truth types. 
We show 9 discovered types in Table \ref{tab:column}. 
\system discovers not only coarse-grained types such as ``year'', ``gender'', or
``state'', but also non-trivial types such as baseball in-game events and central European
cities (last two types). It is quite hard to obtain these fine-grained types
in the multi-class classification setting. \system makes them accessible by
nearest neighbor queries (e.g., find all relevant columns to a given column)
which are useful in data discovery applications.
\system achieves this result by using only 1k column pair labels instead of 
more than 80k multi-class single-column labels required by previous column type detection solutions~\cite{DBLP:conf/kdd/HulsebosHBZSKDH19,DBLP:journals/pvldb/ZhangSLHDT20}.

\subsection{More details about the data cleaning Experiment}\label{app:dc}

\begin{table}[!ht]
\caption{Statistics of correction candidates of Baran vs. \system. ``FT'' and ``no FT'' indicate whether the candidates are
fine-tuned on the labeled tuples or not. \system uses
candidates of slightly lower coverage and larger sizes.}
\label{tab:cleaning_stat}
\scriptsize
\centering
\begin{tabular}{cccccc} \toprule
                          & beers & hospital & rayyan & tax      & AVG    \\ \midrule
\%coverage (Baran, no FT) & 87.30 & 98.82    & 10.97  & 71.10    & 67.05  \\
\%coverage (Baran, FT)    & 99.59 & 99.02    & 51.69  & 94.21    & 86.13  \\
\%coverage (us)           & 94.87 & 89.49    & 51.41  & 92.66    & 82.11  \\ \midrule
\#cand (Baran, no FT)     & 49.00 & 46.23    & 222.60 & 1,297.25 & 403.77 \\
\#cand (Baran, FT)        & 49.63 & 46.63    & 239.80 & 1,302.26 & 409.58 \\
\#cand (us)               & 63.40 & 68.34    & 215.57 & 1,442.27 & 447.40  \\ \bottomrule
\end{tabular}
\end{table}

\begin{figure*}[ht]
	\centering
	\includegraphics[width=0.98\textwidth]{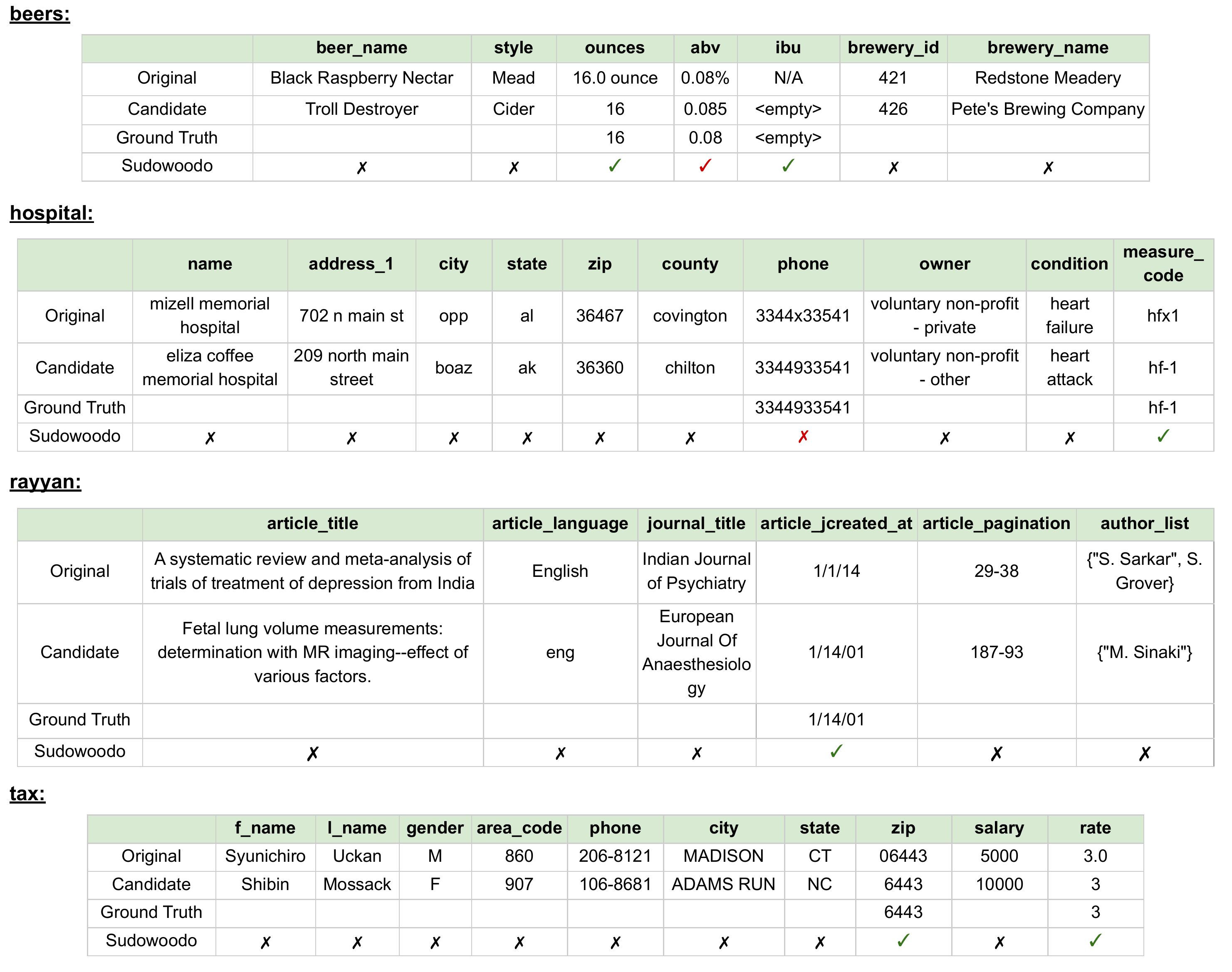}
	\caption{\small Example repairs of \system. 
		The 1st row of each table contains
		a row sampled from the original $\{$beers, hospital, rayyan, tax$\}$ table. The 2nd row contains the top correction
		candidate proposed for each cell. The
		3rd row contains the ground truth corrections if the cell
		is the original erroneous otherwise empty. The last row
		contains the binary predictions from \system of whether
		the candidate correction should be taken or not.}
	\label{fig:dc-examples}
\end{figure*}

As mentioned in Section \ref{sec:experiment}, we follow the
settings of Baran~\cite{DBLP:journals/pvldb/MahdaviA20}
for generating candidate corrections.
Some example candidates can be found in Figure \ref{fig:dc-examples}
for the 4 datasets. The candidate generation process of Baran
consists of a pre-training phase and a fine-tuning phase.
The pre-training phase generates candidates based on 
the dirty table and global knowledge 
such as Wikipedia revision history. The fine-tuning phase 
further updates the pre-trained candidates by adding new candidates
found on the 20 labeled tuples. Since Baran and \system use
different sampling methods to collect the labeled tuples
(active vs. uniform sampling), their
correction candidates are slightly different.
We ensure that \system does not receive extra labeled signals
from the candidate sets by using only the 20 labeled tuples
to fine-tune the pre-trained candidates.

We summarize the difference in Table \ref{tab:cleaning_stat}.
We measure the quality of the candidates by their coverage
(the fraction of error cells have their ground truth correction
in the candidates) and the candidate set size (\#cand).
We notice that without fine-tuning (FT), the quality of the
candidates is quite low. For example, they cover 10.97\% of
errors in the rayyan dataset. After fine-tuning, 
the candidates have much better quality in terms of coverage,
but the candidate set size can still be quite large, e.g., $>$1000
candidates for the tax dataset.
We note that candidates for \system are worse than those
for Baran (lower coverage and larger \#cand), but \system still
achieves a better overall correction F1 score compared to Baran (Table \ref{tab:dc}).

\smallskip
\noindent
\textbf{Ablation analysis on data cleaning. }
We also conduct an ablation analysis on the data cleaning datasets
similar to EM.
Table \ref{tab:cleaning_ablation} summarizes the results.
The performance of \system is less sensitive when optimizations
are removed in data cleaning tasks compared to EM datasets.
\system without the redundancy regularization (RR)
achieves the best overall results. The other 2 optimizations,
cutoff data augmentation (cutoff) and 
clustering-based negative sampling (cls), are most effective
on the beers and hospital datasets, respectively.

\begin{table}[!ht]
\caption{Ablation analysis on data cleaning datasets.}
\label{tab:cleaning_ablation}
\centering
\begin{tabular}{cccccc} \toprule
                          & beers & hospital & rayyan & tax   & AVG   \\ \midrule
Baran + perfect ED & 90.43 &	90.48 &	63.51	& 80.73 & 81.29 \\ \midrule
Sudowoodo (-cutoff)       & 90.47 & 88.89    & 63.53  & 90.95 & 83.46 \\
Sudowoodo (-RR)           & 91.48 & 89.20    & 63.75  & 91.06 & 83.87 \\
Sudowoodo (-cls)          & 91.44 & 88.04    & 62.72  & 90.40 & 83.15 \\
Sudowoodo (-cls, -cutoff) & 91.60 & 88.26    & 62.29  & 90.93 & 83.27 \\
Sudowoodo (-cutoff, -RR)  & 91.16 & 88.86    & 62.68  & 90.97 & 83.42 \\
Sudowoodo (full)          & 90.71 & 89.01    & 62.91  & 91.22 & 83.46 \\ \bottomrule
\end{tabular}
\end{table}

\begin{table*}[h!t]
\centering
	\scriptsize
	\caption{Performance gain of \system over Ditto under different
		difficulty levels (top half). We split the test sets into 5 subsets (1-5)
		of the same size (also same positive ratio).
		We use the Jaccard similarity of positive/negative pairs
		to measure the difficulty of each split.
		Each split contains positive/negative pairs with Jaccard
		within
		a certain range (bottom half). A split is harder if it contains positive pairs
		of lower Jaccard similarity and negative pairs of higher Jaccard similarity, meaning that a method relying on textual syntactic similarity (superficial representations) is less likely to achieve
		high F1. \system outperforms Ditto across different difficulty
		levels especially on level 1 and 5.
	}\label{tab:profiling}
	\begin{tabular}{ccccccccccc} \toprule
		& \multicolumn{2}{c}{Abt-Buy}         & \multicolumn{2}{c}{Amazon-Google}   & \multicolumn{2}{c}{DBLP-ACM}        & \multicolumn{2}{c}{DBLP-Scholar}    & \multicolumn{2}{c}{Walmart-Amazon}  \\ \midrule
		Difficult & Ditto            & Sudowoodo        & Ditto            & Sudowoodo        & Ditto            & Sudowoodo        & Ditto            & Sudowoodo        & Ditto            & Sudowoodo        \\ \midrule
		5         & 65.88            & 81.82 \green{(×1.24)}    & 13.04            & 19.78 \green{(×1.52)}    & 73.08            & 82.29 \green{(×1.13)}    & 54.05            & 63.22 \green{(×1.17)}    & 20.83            & 35.40 \green{(×1.7)}     \\
		4         & 66.67            & 70.21 \green{(×1.05)}    & 45.07            & 56.57 \green{(×1.26)}    & 98.27            & 99.43 \green{(×1.01)}    & 88.14            & 95.75 \green{(×1.09)}    & 22.22            & 53.16 \green{(×2.39)}    \\
		3         & 74.47            & 81.72 \green{(×1.1)}     & 57.97            & 80.90 \green{(×1.4)}     & 100.00           & 100.00 \green{(×1)}      & 93.57            & 97.90 \green{(×1.05)}    & 33.96            & 72.94 \green{(×2.15)}    \\
		2         & 71.60            & 86.75 \green{(×1.21)}    & 64.79            & 91.67 \green{(×1.41)}    & 99.43            & 100.00 \green{(×1.01)}   & 95.30            & 100.00 \green{(×1.05)}   & 25.53            & 84.62 \green{(×3.31)}    \\
		1         & 68.35            & 91.76 \green{(×1.34)}    & 75.68            & 97.78 \green{(×1.29)}    & 98.85            & 99.43 \green{(×1.01)}    & 99.77            & 99.53 \green{(×1)}       & 22.22            & 76.92 \green{(×3.46)}    \\
		\midrule
		Difficult & pos              & neg              & pos              & neg              & pos              & neg              & pos              & neg              & pos              & neg              \\
		\midrule
		5         & {[}0.11, 0.23{]} & {[}0.29, 0.56{]} & {[}0.31, 0.52{]} & {[}0.54, 0.86{]} & {[}0.21, 0.71{]} & {[}0.34, 0.94{]} & {[}0.24, 0.53{]} & {[}0.41, 0.85{]} & {[}0.26, 0.49{]} & {[}0.55, 0.74{]} \\
		4         & {[}0.23, 0.26{]} & {[}0.24, 0.29{]} & {[}0.52, 0.59{]} & {[}0.48, 0.54{]} & {[}0.71, 0.78{]} & {[}0.29, 0.34{]} & {[}0.53, 0.61{]} & {[}0.35, 0.41{]} & {[}0.50, 0.55{]} & {[}0.50, 0.55{]} \\
		3         & {[}0.27, 0.32{]} & {[}0.21, 0.24{]} & {[}0.59, 0.67{]} & {[}0.44, 0.48{]} & {[}0.78, 0.81{]} & {[}0.26, 0.29{]} & {[}0.61, 0.68{]} & {[}0.32, 0.35{]} & {[}0.56, 0.62{]} & {[}0.47, 0.50{]} \\
		2         & {[}0.32, 0.37{]} & {[}0.18, 0.21{]} & {[}0.67, 0.73{]} & {[}0.39, 0.44{]} & {[}0.82, 0.86{]} & {[}0.23, 0.26{]} & {[}0.68, 0.79{]} & {[}0.29, 0.32{]} & {[}0.62, 0.68{]} & {[}0.44, 0.47{]} \\
		1         & {[}0.37, 0.54{]} & {[}0.10, 0.18{]} & {[}0.74, 0.91{]} & {[}0.27, 0.39{]} & {[}0.86, 0.97{]} & {[}0.14, 0.23{]} & {[}0.79, 0.95{]} & {[}0.17, 0.29{]} & {[}0.68, 0.87{]} & {[}0.33, 0.44{]}
		\\ \bottomrule
\end{tabular}\end{table*}

\begin{figure*}[ht]
	\centering
	\includegraphics[width=0.98\textwidth]{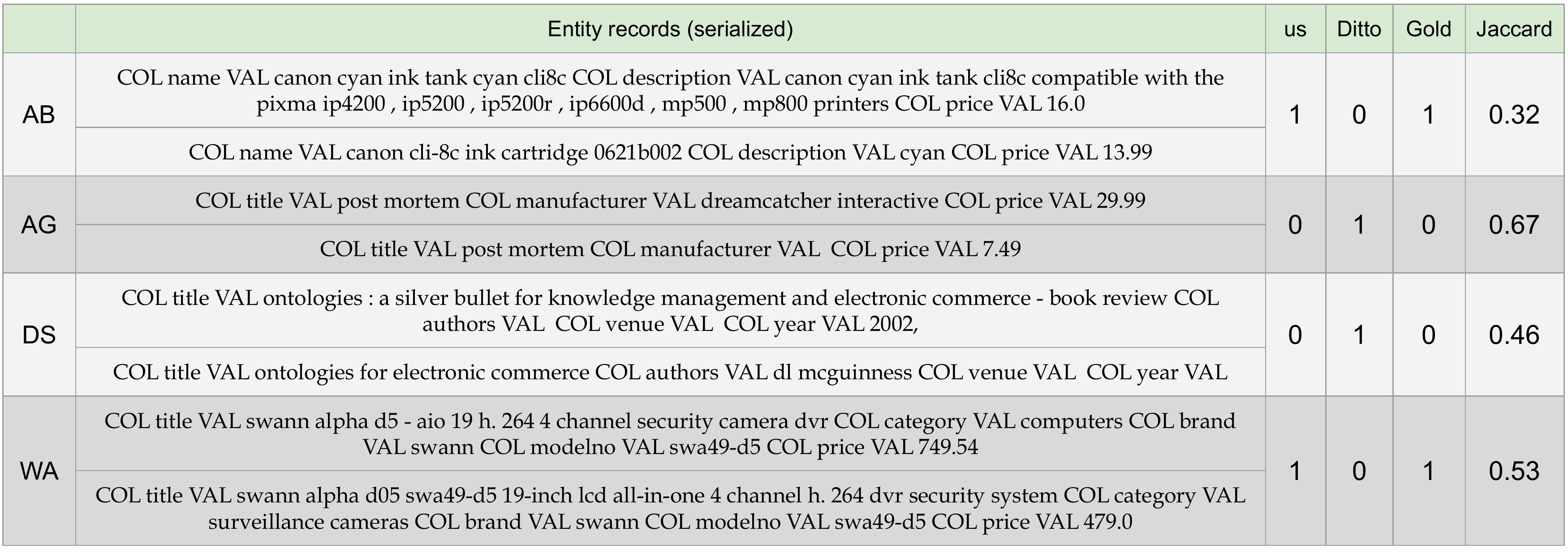}
	\caption{Example pairs in test sets where \system predicts correctly but Ditto does not.}
	\label{fig:em-examples}
\end{figure*}

\subsection{Data profiling and error analysis}\label{app:case}

\subsubsection{Data profiling on Entity Matching}
\label{app:profiling}

We conduct a data profiling analysis on the 5 datasets
used for the evaluation of semi-supervised EM.
The goal of the analysis is to understand in what EM data
\system has performance advantages over previous methods.
We conduct the analysis by splitting each test set into 
5 ``difficulty'' levels. We measure the difficulty of a subset
by the \emph{average Jaccard similarity} of the positive and
negative classes. 
Formally, given a pair of entities $(e, e')$
in the test set, the Jaccard similarity is computed as
$$ |\mathsf{tokens}(e) \cap \mathsf{tokens}(e')| \ / \
|\mathsf{tokens}(e) \cup \mathsf{tokens}(e')|$$
where $\mathsf{tokens}(e)$ is the set of tokens appear
in entity $e$.

Jaccard similarity is widely used in rule-based methods
for measuring textual or syntactic similarity between two
text sequences. Two entity records are likely to match if they
have high Jaccard similarity.
Intuitively,
a given EM dataset $D$ is ``easy'' if the positive class of $D$
has high average Jaccard similarity and the negative class
has low average Jaccard similarity because a classifier
based on textual or syntactic similarity can likely separate
the two classes. On the other hand, a dataset is ``hard''
if the positive/negative class has low/high Jaccard similarity
making them harder to be separated.

Following the above idea, we create the 5 difficulty levels 
of each dataset by splitting them into 5 subsets of equal sizes
and equal positive ratios. We sort the entity pairs by their
Jaccard similarity such that level 5 (the hardest) has
the lowest similarity among the positive class and 
the highest similarity among the negative class.
Level 1 (the easiest) is the exact opposite.
The bottom of Table \ref{tab:profiling} shows the 
ranges of Jaccard similarity of each split of each dataset.
For example, DA is the easiest dataset 
since the hardest level still has high/low Jaccard similarity 
in the positive/negative class.

The top of Table \ref{tab:profiling} shows the performance
gain of \system over Ditto, the SOTA DL-based entity matching 
method, under different difficulty levels.
We notice that \system outperforms Ditto across different
difficulty levels. \system also performs relatively well 
and more consistently when we increase the difficulty levels.
For example, \system significantly outperforms Ditto 
at the hardest level of the Abt-Buy dataset (1.24x),
the Amazon-Google dataset (1.52x), and the DBLP-Scholar dataset (1.17x).
The Walmart-Amazon dataset is an exception where the Ditto
model fails to obtain a sufficient F1 score in all cases.
The results indicate the necessity of high-quality representations
to achieve high F1 scores for harder EM tasks.

We further verify the result on example pairs where
\system predicts correctly but Ditto does not 
(Figure \ref{fig:em-examples}).
For the example pair from the AB dataset,
the Jaccard similarity is relatively low as Ditto predicts
it as a non-match, but \system correctly predicts it as a match
likely by capturing the matching product ID ``cli8c'' which appears
in both entries.
For the AG dataset example of high Jaccard similarity, \system correctly predicts it
as non-match likely by capturing the price difference.

\subsubsection{Data Cleaning error analysis}

In Figure \ref{fig:dc-examples}, we conduct 
an error analysis on 4 rows sampled from each
data cleaning dataset. Each table contains the original
row (Original), the top candidate corrections proposed by
Baran's candidate generator (Candidate), 
the ground truth correction (Ground Truth) if the cell
is erroneous, and \system's predictions of whether the 
candidate should be taken or not.

Notice that the correction candidates can be quite confusing, e.g., 
replacing language ``English'' with ``eng'', repairing 
US state ``al'' with ``ak'', or ``heart failure'' with 
``heart attack''. All these candidates seem to be valid but
the original cell is clean so repairs are not necessary.
\system predicts all these cases correctly.

Among the cells where corrections are needed, 
\system predicts most cases correctly (5/7). The first mistake
of correcting an abv ``0.08\%'' into ``0.085'' is due to
the ground truth correction not present in the candidate set.
The second mistake of not correcting ``3344x33541''
into ``3344933541'' (replacing ``x'' with ``9'') 
is due to multiple valid replacements exist 
(any digits from 0 to 9). While candidate generation is 
currently not in the \system pipeline, 
a promising future work 
can be to couple \system with a pre-trained sequence generation model
such as GPT-2 to generate and select high-quality corrections
in an end-to-end manner.



\subsection{Fully-supervised entity matching}

\setlength{\tabcolsep}{2.5pt}
\begin{table}[!ht]
\caption{Statistics of EM datasets.  }\label{tab:em_dataset_full}
\small
\begin{tabular}{cccccc} \toprule
Datasets            & TableA & TableB & Train+Valid & Test & \%pos   \\ \midrule
Abt-Buy (AB)        & 1,081   & 1,092   & 7,659        & 1,916 & 10.74\% \\
Amazon-Google (AG)  & 1,363   & 3,226   & 9,167        & 2,293 & 10.18\% \\
Beer                & 4,345      & 3,000      & 359         & 91   & 15.11\% \\
DBLP-ACM (DA)       & 2,616   & 2,294   & 9,890        & 2,473 & 17.96\% \\
DBLP-Scholar (DS)   & 2,616   & 64,263  & 22,965       & 5,742 & 18.63\% \\
Fodors-Zagats       & 533      & 331      & 757         & 189  & 11.63\% \\
iTunes-Amazon       & 6,906      & 55,923      & 430         & 109  & 24.49\% \\
Walmart-Amazon (WA) & 2,554   & 22,074  & 8,193        & 2,049 & 9.39\% \\ \bottomrule
\end{tabular}
\end{table}

\begin{table}[ht]
	\small
	\caption{F1 scores for fully-supervised EM.}\label{tab:super}
	\resizebox{0.48\textwidth}{!}{
	\begin{tabular}{ccccc}
		\toprule
		 & \dm & \ditto & \system (w/o \textsf{RR})& \system  \\
		\midrule
Abt-Buy            & 62.8 & 89.8 & 98.3 \green{(+8.5)}  & 98.3 \green{(+8.5)}  \\
Amazon-Google      & 70.7 & 75.6 & 92.9 \green{(+17.3)} & 94.5 \green{(+18.9)} \\
Beer               & 78.8 & 94.4 & 100.0 \green{(+5.6)} & 96.6 \green{(+2.2)} \\
DBLP-ACM           & 98.5 & 99.0   & 100.0 \green{(+1.0)} & 100.0 \green{(+1.0)} \\
DBLP-Scholar & 94.7 & 95.6 & 98.9 \green{(+3.3)}  & 98.9 \green{(+3.3)}  \\
Fodors-Zagats        & 100.0  & 100.0  & 100.0 \green{(+0.0)}  & 100.0 \green{(+0.0)}  \\
iTunes-Amazon      & 91.2 & 97.1 & 98.2 \green{(+1.1)}  & 98.1 \green{(+1.0)}  \\
Walmart-Amazon     & 73.6 & 86.8 & 95.7 \green{(+8.9)}  & 93.3 \green{(+6.5)} \\
		\bottomrule
	\end{tabular}}
\end{table}

Besides the semi-supervised setting, \system also performs well on the fully-supervised setting for the EM task. For this experiment, we include three additional datasets Beer, Fodors-Zagats, and iTunes-Amazon, where the number of labeled data is much less than the datasets used in the semi-supervised learning. The statistics of the datasets are shown in table~\ref{tab:em_dataset_full}.

We set the learning rate of \system to 3e-5 and the number of fine-tuning epochs to 40. 
Since all training examples are used, there is no need to apply the pseudo-labeling step.
We apply each DA operator listed in Table \ref{tab:defaultda} to find the optimal choice for each dataset. 
In addition to the full-fledged \system, we also report the results of \system (w/o \textsf{RR})
which removes the Redundancy Regularization optimization in Section~\ref{sec:barlow_twins} from \system.
Note that one potential optimization for \system is to
apply its blocking stage to further reduce the size of the train/test
sets which will naturally increase the F1 score. 
We exclude this optimization to make the comparison more fair
(same in semi-supervised and unsupervised training).

We summarize the results in Table \ref{tab:super}.
\system outperforms the previous SOTA matching solution \ditto by up to 18.9\% F1 score
and 5.2\% on average. In 3 out of the 8 tasks, \system (w/o \textsf{RR}) achieves the perfect 100\% F1 score.
The performance gain of \system is mainly due to its capability of learning high-quality entity representations
and the similarity-aware fine-tuning steps. 
We also find that the Redundancy Regularization optimization is useful in situations where the task is 
relatively more difficult, such as on the Amazon-Google dataset.

\end{document}